# Core–Shell Confinement Blocks Hydride Formation: The Impact of Surface Oxides on Hydrogen Sorption in Nanoporous FeTi

**Lukas Schweiger**[a*], **Florian Spieckermann**[a*], Michael Burtscher[a], Stefan Wurster[b], Sebastian Stock[c], Nikolaos Kostoglou[d,a], Oskar Paris[c], Alexander Schökel[e], Fahim Karimi[f], Gökhan Gizer[f], Claudio Pistidda[f], Daniel Kiener[a], Jürgen Eckert[a,b]

a Department of Materials Science, Montanuniversität Leoben, Franz Josef-Straße 18, 8700 Leoben, Austria

b Erich Schmid Institute of Materials Science, Austrian Academy of Sciences, Jahnstraße 12, 8700 Leoben, Austria

c Chair of Physics, Montanuniversität Leoben, Franz Josef-Straße 18, 8700 Leoben, Austria

d Institute of Geoenergy, Foundation for Research and Technology – Hellas, 73100 Chania, Greece

e Deutsches Elektronen-Synchrotron DESY, Notkestraße 85, 22607 Hamburg, Germany

f Department of Materials Design, Institute of Hydrogen Technology, Helmholtz-Zentrum Hereon, Max-Planck-Straße 1, 21502 Geesthacht, Germany

*Corresponding authors: lukas.schweiger@unileoben.ac.at, florian.spieckermann@unileoben.ac.at

# Abstract

Metal hydrides remain an intriguing alternative to conventional gaseous and liquid hydrogen storage methods, offering high volumetric storage density and enhanced hydrogen storage safety at ambient conditions. In this regard, the intermetallic compound FeTi is one of the most promising storage materials. However, its widespread industrial application remains challenging due to the need for activation, slow initial kinetics, large hysteresis, and high material costs.

In this study, we aim to overcome these limitations by devising an alternative synthesis pathway to prepare nanoporous and ultra-fine porous FeTi with controlled grain and ligament sizes, allowing us to study the obtained well-defined microstructures in detail. In particular, we observe the confinement of the FeTi phase by surface oxides, which can be correlated with the hydrogen sorption properties of the respective material. These experimental results are further supported by an analytical model allowing the calculation of the absorption pressure as a function of microstructure-dependent elastic stresses. Additionally, we show that such stresses also influence the absorption-desorption hysteresis.

This study lays the groundwork for the controlled and systematic study of the processing-structure-properties relations in metal hydrides and FeTi in particular, thereby paving the way to cost-effective and efficient hydrogen storage solutions based on metal hydrides.

# Introduction

Hydrogen is regarded as an essential energy carrier for a climate-neutral economy,[1–6] in which its production is possible from renewable energy sources via water electrolysis[7] or methane pyrolysis[8]. The green gas can be employed in various ways, for example, in fuel cells, internal combustion engines, as a feedstock for the chemical industry, or for the production of green steel.[2,9] However, gaseous hydrogen has a low volumetric energy density, and therefore, its ubiquitous usage is strongly connected to the availability of efficient and cheap storage solutions with high energy densities.[5,10]

Many storage options for hydrogen exist, such as pressurized (up to 700 bar) and liquified (20 K) hydrogen, physisorption on nanoporous and high-surface area materials, or the intermittent synthesis of ammonia.[11,12] Another way of achieving high volumetric storage densities is the use of metal hydrides.[13,14] These metal hydrides form by the reversible absorption of hydrogen in metals with high hydrogen affinity, and by tuning this affinity, the properties of the storage medium can be controlled and adjusted to certain specifications.[15] Additional advantages of metal hydrides are the potential for lower storage pressures and near-ambient operating temperatures, as well as high safety, efficiency, and cyclability.[10,13]

A promising metal hydride is the intermetallic compound FeTi, which exhibits favorable thermodynamic and kinetic properties, along with a gravimetric hydrogen capacity of 1.9 wt.%.[16–19] In particular, hydrogen absorption can occur at moderate pressures (< 30 bar) – which can be directly provided by electrolyzers – while desorption can be achieved using the waste heat of fuel cells.[13,20,21] However, the material decrepitates upon repeated absorption/desorption, i.e., it undergoes fragmentation, thus limiting its cyclability and performance.[22] Moreover, the main drawback of FeTi is the need for an activation treatment at elevated temperatures and hydrogen pressures before absorption[23], coupled with its easy deactivation when in contact with $H_2O$, $O_2$, or CO.[24] Approaches for mitigating the (de-)activation problem include alloying additions[25] like Mn[26] and mechanical deformation using, e.g., cold rolling[27,28], ball milling[29–31] or high-pressure torsion (HPT)[32,33].

Promising results of HPT-deformed FeTi indicated absorption without any activation treatment and no deactivation in air.[32] This improved absorption behavior was traced to the induced segregation of Fe-rich precipitates on the surface, microcracks, and lattice defects (dislocations).[34] Overall, HPT is very versatile for synthesizing metastable materials, supersaturated solid solutions, nanocrystalline materials, and nanocomposites with superior mechanical and functional properties.[35–39] If one of the constituent phases is selectively dissolved, these nanocomposites can also serve as templates for the design of nanoporous metallic materials.[40–43] Such nanoporous structures have large surface areas as well as high interface and defect densities beneficial for activation, but also internal free volume that could mitigate decrepitation and enhance both dimensional and mechanical stability during absorption and desorption cycles. For example, the rapid kinetics, along with the dimensional stability of the ammonium salt $Mg(NH_3)_6Cl_2$ during ammonia uptake/release (10 wt.%), have been attributed to the formation of nanopores.[44,45] Consequently, nanoporous metal hydrides could play a critical role in ensuring long-term stability and performance, potentially justifying higher initial production and material costs.

Apart from (de)activation and mechanical instability, the high material and production costs are additional challenges for the application of metal hydrides.[10,22] Although FeTi fares better than other materials such as $LaNi_5$, especially when considering its production from scrap,[46] reducing its cost is still an ongoing research effort. One solution would be to synthesize FeTi directly from $FeTiO_3$, i.e., ilmenite. In fact, such production approaches often result in porous structures or fine powders.[47–52] Even though such approaches can be highly promising, they sometimes fall short in achieving the required absorption capacities and activation behavior. Therefore, it is essential to clarify how structural size and morphology—whether particles, ligaments, or pores—influence hydrogen sorption performance, and model systems with tunable structural features are key to gaining this understanding.

To understand and overcome these performance limitations, we synthesized nanoporous FeTi derived from HPT-deformed FeTi−Cu nanocomposites.[53,54] This system serves as a model system for studying the impact of pore and ligament sizes and morphologies on the activation and sorption properties of FeTi. Indeed, this investigation revealed that hydrogen absorption is blocked by a confinement effect in nanoporous FeTi, while larger ligaments do not show prohibited hydride formation. Apart from these insights, porous FeTi derived from severely plastic deformed composites constitutes a promising and highly tunable material class for hydrogen storage and other applications, and could exhibit superior functional properties and mechanical stability facilitated by high surface area, high defect densities, and high strength at low weight.

# Experimental

**Sample preparation**

FeTi was synthesized by arc melting (Arc Melter AM 0.5, Edmund Bühler GmbH). The starting materials were iron flakes (99.99+ % purity; HMW Hauner GmbH) and titanium granules (99.995 % purity; HMW Hauner GmbH) in stoichiometric ratios (50:50 at.%). The arc-melted ingot was crushed and subjected to ball milling with copper powder (99.9 % purity; Alfa Aesar) to achieve an FeTi−Cu powder blend with 35 vol.% (42 wt.%) Cu. This blend was compacted (7.5 GPa, 60 s) at room temperature (RT) under an Ar atmosphere and subsequently deformed by HPT at 400 °C for 50 revolutions (6 GPa, 1.27 rounds per minute). The anvils had a cavity diameter and depth of 8 and 0.15 mm, respectively; the resulting HPT disk had a corresponding diameter of 8 mm and an average thickness of 0.5 mm. Further details on the sample preparation can be found in previous works.[53,54]

The shear strain $\gamma$ introduced at a radius $r$ in the HPT disk was calculated based on[55]

$$\gamma = \frac{2\pi n r}{t} \quad , \tag{1}$$

with $n$ being the number of revolutions, and $t$ the sample thickness. Subsequently, some as-HPT deformed composites were subjected to annealing in a vacuum furnace (Xerion Advanced Heating Ofentechnik GmbH, vacuum tube furnace, type Xtube) at $10^{-6}$ mbar and 700 °C for 1 h.

The FeTi−Cu composites, either as-HPT deformed or annealed, were immersed (up to 48 h) in a stirred 1 M $(NH_4)_2S_2O_8$ solution at RT to selectively dissolve the Cu phase. After etching, the samples were cleaned by subsequent immersion in water, ethanol, isopropanol, and acetone. The procedure of sample preparation, starting from the powders, is illustrated in **Figure 1 (a)**.

**Microstructural and mechanical characterization**

The HPT disks were halved and mirror-polished to allow a cross-sectional analysis of the microstructure using scanning electron microscopy (SEM; LEO type 1525, Carl Zeiss GmbH / Magna, Tescan), including energy dispersive X-ray spectroscopy (EDX; XFlash 6–60, Bruker corporation) and electron backscatter diffraction (EBSD; Bruker e⁻-FlashFS, Bruker corporation). Subsequent microhardness measurements ($HV_{0.2}$) were performed on the same cross-sections (DuraScan, ZwickRoell GmbH). For porous FeTi, the cross-sections were prepared by ion milling (E-3500 Ion Milling System, Hitachi).

Transmission electron microscopy (TEM) images were acquired in axial view using scanning TEM (STEM) and high-resolution TEM (HR-TEM) modes (JEM-2200FS microscope, JEOL Ltd.). Electron-transparent specimens were obtained by polishing, dimple grinding, and ion milling disks previously cut to a 3 mm radius. The porous structure for TEM investigations was prepared by immersing the electron-transparent samples in a 1 M $(NH_4)_2S_2O_8$ solution. The electron transparent specimens were additionally investigated by transmission Kuckuchi diffraction (TKD) using the Optimus TKD detector head (Bruker Nano GmbH) attached to the aforementioned EBSD detector within a SEM (Magna, Tescan)

Low-pressure gas sorption experiments (Autosorb iQ3 gas sorption analyser, Anton Paar QuantaTec) were made using nitrogen ($N_2$) gas of ultra-high purity (99.999 %) as adsorbate. Vacuum degassing was performed at 105 °C for 24 h. Subsequent adsorption and desorption isotherms were measured at 77 K, utilizing a liquid nitrogen dewar vessel, across a relative pressure ($p/p_0$) range of $10^{-3}$ to 0.99.

Synchrotron radiation X-ray diffraction (SR-XRD) measurements were conducted at the P02.1 Powder Diffraction and Total Scattering Beamline of PETRA III (DESY Hamburg) in transmission geometry using a photon energy of 60 keV and a Dectris EIGER2X CdTe 1M-W detector. Calibration was performed by using a $CeO_2$ reference standard, and data processing and integration was carried out with the pyFAI software package.[56] Rietveld refinement of the obtained patterns, including the determination of the coherent scattering domain (CSD) sizes and microstrains, was performed using the software package Profex.[57] A custom-designed high-pressure cell[58] was used for the *in-situ* experiments at elevated temperatures and hydrogen pressures of up to 50 bar of hydrogen. The porous samples were crushed in a glovebox before loading the resulting powders into sapphire capillaries with an inner diameter of 0.8 mm.

Ultra-Small-Angle Neutron Scattering (USANS) and Small-Angle Neutron Scattering (SANS) measurements were conducted at Australia's Nuclear Science and Technology Organisation (ANSTO) using the Kookaburra USANS and Quokka SANS instruments, respectively.[59,60] USANS measurements employed a 29 mm aperture and 4.74 Å neutrons (Si(111)) covering a $q$ range of $4 \cdot 10^{-5}$ to $8 \cdot 10^{-4}$ Å$^{-1}$, overlapping with the SANS data. The Quokka instrument used a 1 m² position-sensitive detector with a neutron velocity selector providing tunable wavelengths from 4.5 to 40 Å, calibrated with a polymethyl methacrylate (PMMA) standard. The SANS data was collected at sample-to-detector distances of 1.3 m, 12 m, and 20 m (with $MgF_2$ lenses) using wavelengths of 5 Å and 8.1 Å for optimal $q$ range coverage.

**Functional characterization**

Hydrogen absorption and desorption properties were characterized using the pressure-composition isotherm (PCI) method with a commercial manometric gas sorption apparatus (iSorb HP1-100, Anton Paar QuantaTec) and ultra-pure (99.999 %) hydrogen gas. The as-etched HPT disks were mechanically fractured, and the resulting pieces were immediately loaded into the sample holder and transferred to the PCI apparatus for hydrogen sorption measurements. The samples were subjected to degassing (at 100°C) and/or activation (at 400°C) before recording PCI curves; details of the corresponding procedures are provided in the Supporting Information. The analysis was done in a temperature range of 45 °C to 80 °C and at pressures of up to 100 bar. The entropy and enthalpy of formation/ decomposition were determined from equilibrium data by using the Van't Hoff equation,

$$\ln\left(\frac{p_{eq}}{p_0}\right) = \frac{\Delta H}{RT} - \frac{\Delta S}{R} \quad , \tag{2}$$

with $p_{eq}$ being the equilibrium plateau pressure in Pa, $p_0$ the standard pressure with a value of 0.1 MPa, $\Delta H$ the enthalpy of hydride formation/decomposition, $\Delta S$ the entropy of hydride formation/decomposition, R the ideal gas constant, and $T$ the temperature in K. Enthalpies of hydride formation, derived from the absorption branch, are reported as negative, whereas enthalpies of hydride decomposition, derived from the desorption branch, are positive.

# Results

**Structural characterization of the FeTi−Cu composites**

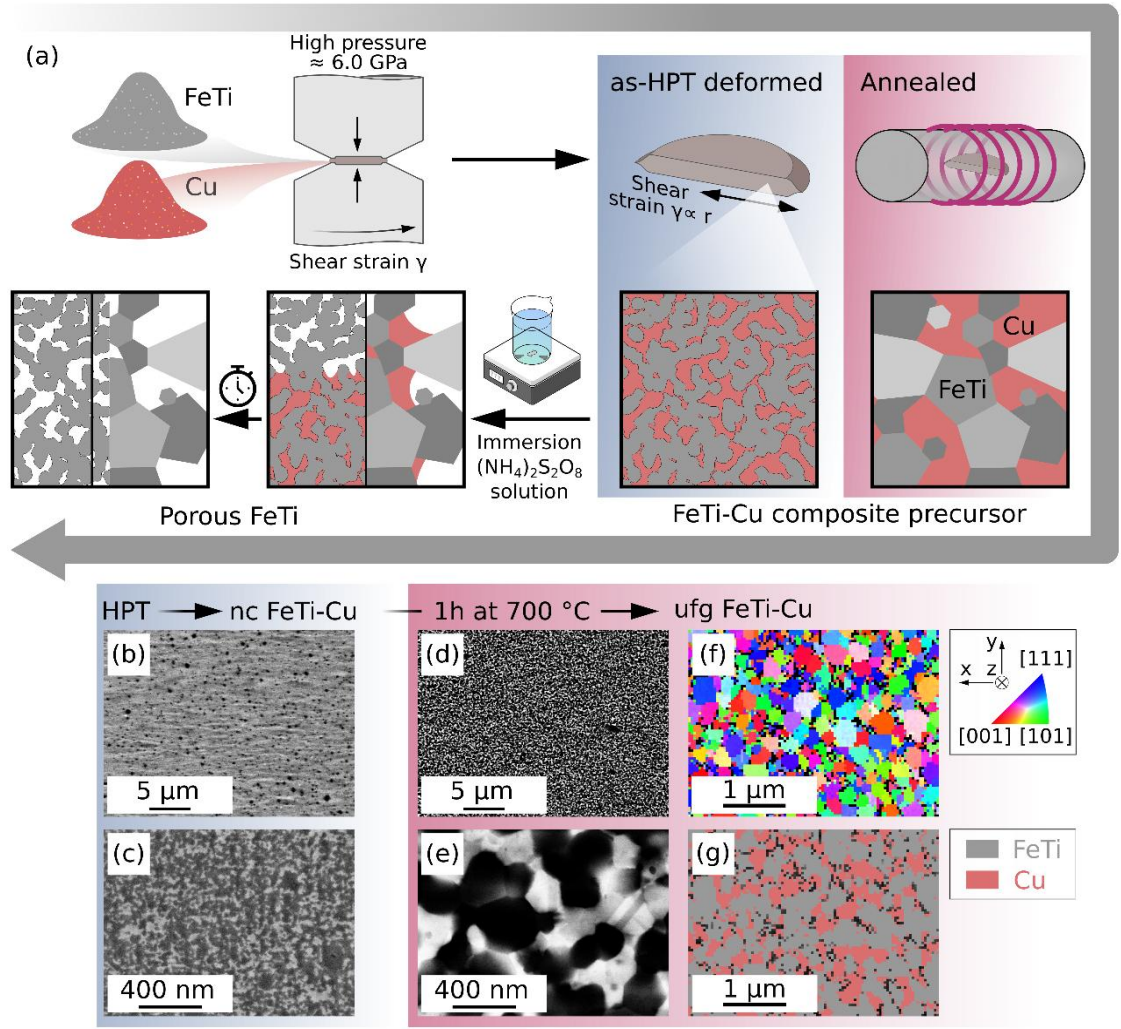

**Figure 1:** (a) Schematic material fabrication process. (b,c) Backscatter electron (BSE) SEM micrographs of as-HPT deformed FeTi−Cu composites with a nanocrystalline (nc) structure and (d,e) subsequently annealed FeTi−Cu composites with an ultra-fine grained (ufg) microstructure; (c) was made with an In-Lens detector. (f,g) Inverse pole figure (IPF) and phase maps obtained via EBSD from annealed composites. All images were taken in tangential direction at $r \approx 3$ mm (nominal shear strain $\gamma \approx 1900$).

The results of the microstructural characterization of the FeTi−Cu composites are provided in **Figure 1**. For details on the microstructure evolution of the FeTi−Cu composite during HPT deformation, the reader is referred to previous studies, which show that high-temperature deformation is essential for obtaining homogeneous nanocomposites.[53,54]

The SEM micrographs in **Figure 1** depict the microstructures of FeTi−Cu composites (b, c) as-HPT deformed and (d-g) subsequently annealed for 1 h at 700 °C. The as-HPT deformed materials exhibited a nanocrystalline (nc) structure, and annealing of these FeTi−Cu nanocomposites resulted in a significant yet controlled coarsening of the composite to an ultra-fine grained (ufg) structure. Consequently, these material states will be referred to as nc (i.e., as-HPT deformed) and ufg (i.e., annealed) FeTi−Cu composites. As shown in **Figures S1-S2** (**Supporting Information**), FeTi and Cu particles exhibit log-normal distributions in both nc and ufg FeTi−Cu composites, with mean FeTi particle sizes of 54.8±0.7 nm and 322±8 nm at $r \approx 3$ mm ($\gamma \approx 1900$). Detailed particle and grain sizes are given in **Table S1** (**Supporting Information**).

Inverse pole figure (IPF) and phase maps derived from EBSD measurements on the ufg FeTi−Cu composite are provided in **Figure 1 (f,g)** and are in line with the other SEM micrographs. These EBSD results, together with the transmission Kuchuchi diffraction (TKD) results further below, reveal negligible texture for both nc and ufg FeTi−Cu composites. The corresponding data is given in **Figures S3-S5** (**Supporting Information**). Consequently, the obtained composite materials are highly isotropic. Although the introduced shear strain during HPT is dependent on the radial position, high enough strains can lead to saturation and a steady-state microstructure.[61–63] The results of radially resolved measurements are given in **Figure S6**, including the chemical, structural, and mechanical characteristics of the composites. Both nc and ufg FeTi−Cu composites exhibit a saturated microstructure at $r \geq 1$ mm, about 94 % of the sample volume. Associated EDX maps are provided in **Figures S7-S8** (**Supporting Information**). Therefore, the saturated structure will dominate the sorption properties of the derived porous materials, and the radial strain gradient can be neglected.

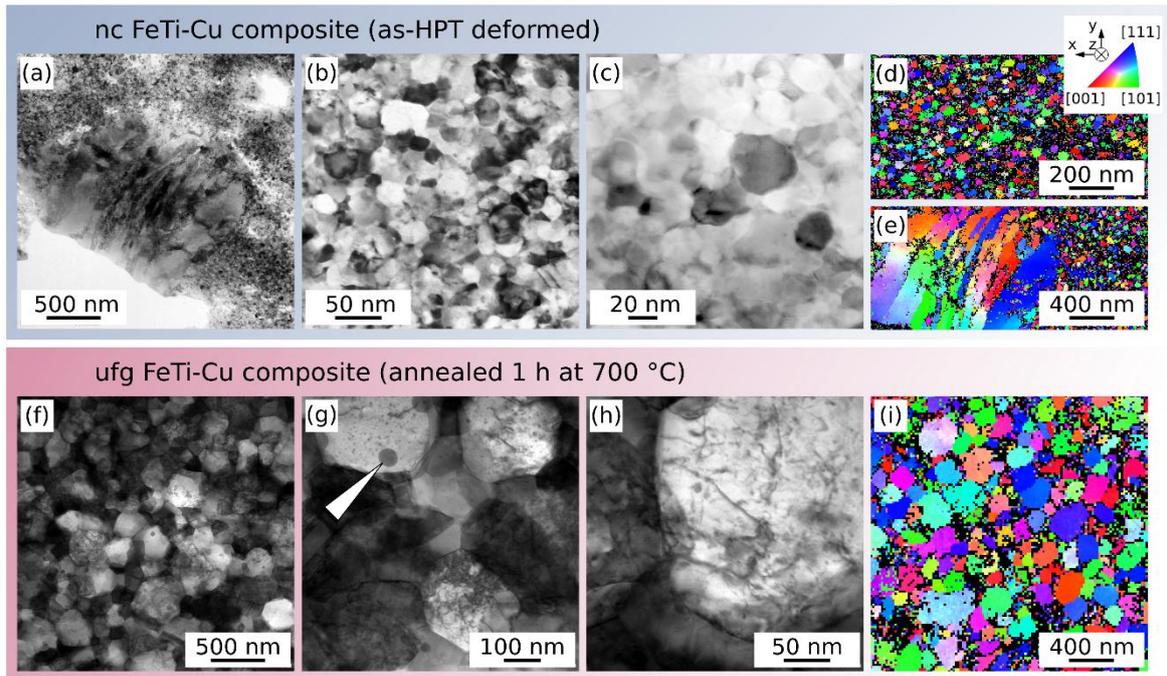

**Figure 2:** (a-c) TEM micrographs of nc FeTi−Cu composites with (d,e) the corresponding IPF map obtained by TKD. (f-h) TEM micrographs of ufg FeTi−Cu composites with (i) the corresponding IPF maps obtained by TKD. The arrow in (g) highlights exemplary precipitates within the ufg material.

**Figure 2** shows the characterization of nc and ufg FeTi−Cu composites using TEM and TKD. **Figure 2 (a)** shows the microstructure of a nc FeTi−Cu composite, with a larger FeTi particle in the center. Despite such heterogeneities, **Figure 2 (b,c)** highlights the otherwise uniform nanocrystalline nature of the composite. IPF maps in **Figures 2 (d,e)** resolve both nanocrystalline FeTi (B2) and Cu (fcc); **Figure 2 (e)** reveals the substructure in the larger FeTi particle shown in **(a)**. **Figure 2 (f-i)** depicts the TEM and TKD results of an ufg FeTi−Cu composite, where, in line with the SEM results, the microstructure exhibited a significantly coarser structure. An arrow in **Figure 2 (g)** also highlights some nanoscale precipitates, visible in both the FeTi and Cu phases. These were identified as $Fe_2Ti$ (P 63/m m c (Nr. 194), see **Figure S9 (Supporting Information**) and $Ti_2Fe$ (F d -3 m (Nr. 227); or the structurally similar $Ti_4Fe_2O$), see **Figure S10 (Supporting Information**), aligning with the SR-XRD results in

**Figure 3**. **Figure 2 (i)** provides an IPF map of an ufg FeTi−Cu composite, showing the significantly coarser grains compared to the nc composites.

The grain sizes determined from TKD for nc (FeTi: 33.7 ± 0.8 nm) and ufg (FeTi: 226 ± 8 nm) FeTi−Cu composites are included in **Figure S6 (e) (Supporting Information)**, and detailed results are provided in **Figures S4-S5** (**Supporting Information**). TKD confirms, in line with EBSD, that both nc and ufg FeTi−Cu composites have insignificant crystallographic texture. TEM-derived EDX maps in **Figures S11-S13** (**Supporting Information**) show distinct Cu-rich and Cu-lean regions in both nc and ufg FeTi−Cu composites. The quantified elemental compositions are plotted in **Figure S14** (**Supporting Information**) and reveal two distinct clusters corresponding to the Cu and FeTi phases, respectively. The ufg composite exhibits an additional cluster associated with $Fe_2Ti$; no cluster related to $Ti_2Fe$ could be found. This chemical partitioning additionally confirms the well-separated two-phase material state in both the as-HPT deformed and annealed FeTi−Cu composites.

# Structural characterization of porous FeTi

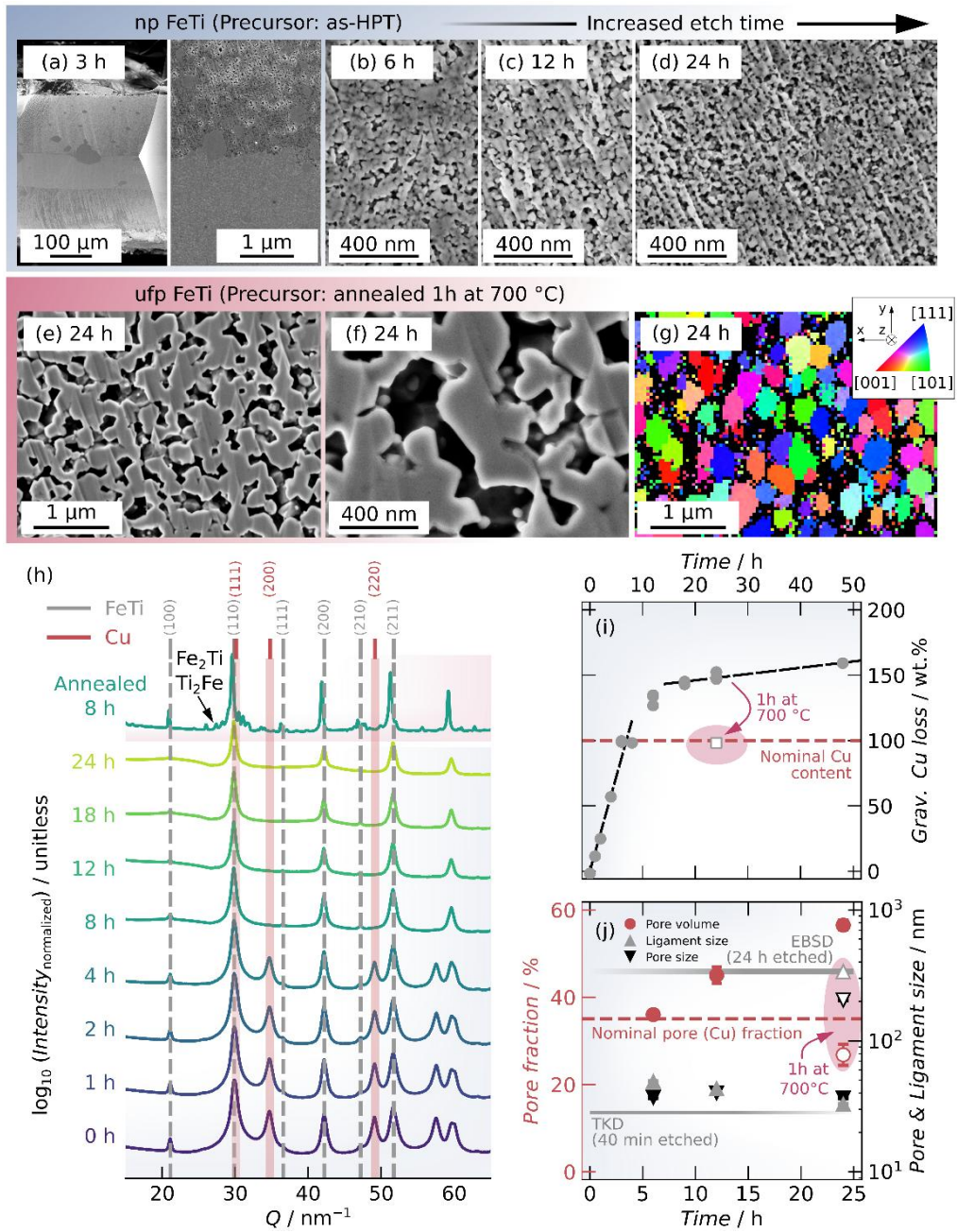

**Figure 3:** SEM micrographs of cross-sections of nc FeTi−Cu composites etched for (a) 3, (b) 6, (c) 12, and (d) 24 h. (e,f) Annealed ufg composites etched for 24 h and (g) the associated IPF map. (h) SR-XRD patterns of nc FeTi−Cu composites etched for different durations, and an ufg FeTi−Cu composite etched for 8 h. Note the logarithmic intensity scaling. (i) Mass loss during etching, and (j) the pore fractions and feature sizes from image analysis. Open symbols represent the values obtained from ufg FeTi−Cu composites and are highlighted by arrows.

SEM micrographs of nc FeTi−Cu composites etched for different durations (up to 48 h) are given in **Figure 3 (a-d)**. Immersion for 3 h yields incomplete Cu dissolution, as indicated by the Cu-rich composite in the disk center, bordered by a well-defined etch front. No Cu-containing core was detected when immersed for 6 h or longer. The related EDX maps confirming this are given in **Figures S15-S16 (Supporting Information)**.

Micrographs of ufg FeTi−Cu composites after etching for 24 h in **Figures 3 (e,f)** show the formation of significantly larger pores and ligaments. These ligaments exhibit a polygonal morphology, as opposed to the pronounced curvature in the nanoporous material. Comparison with **Figure 1** shows that this reflects the microstructure of the ufg FeTi−Cu composite, but also suggests a negligible attack of the etchant on the FeTi ligaments. The EBSD-derived IPF map in **Figure 3 (g)** of the same ion-polished cross-section confirms that the ligaments consist of FeTi, with most ligaments consisting on average of one FeTi grain.

The SR-XRD results for nc FeTi−Cu etched for different durations and measured in transmission are given in **Figure 3 (h)**. The intensity, normalized to the maximum peak intensity of the respective pattern, is plotted against the scattering vector $Q$. The initial composite consists of two phases, Cu and FeTi, but with continued etching, the Cu peaks become weaker and eventually disappear. Quantification by Rietveld refinement confirmed complete Cu dissolution after 8 h, in line with the gravimetric analysis of the etching process as shown in **Figure 3 (i)** and the SEM analysis presented in **Figure 3 (a-d)**. The diffraction patterns of the ufg FeTi−Cu composites exhibited significantly sharper peaks, and after 8 h of etching, no Cu peaks could be detected anymore. Detailed results of the Rietveld refinement, including lattice constant, microstrains, and coherent scattering domain (CSD) sizes, are given in **Figure S17** (**Supporting Information**).

**Figure 3 (i)** plots the mass loss during etching, revealing two distinct etch stages for the nc FeTi−Cu composite. Initially, the mass loss is rapid due to the dissolution of the metallic Cu. Upon removal of Cu, the etch rate decreases significantly, though it remains non-zero, likely due to continued etching of the FeTi ligaments. Planned interval tests, as shown in **Figure S18 (Supporting Information)**, confirm that the variation in etching rates is related to the material. On the contrary, the gravimetric analysis of the ufg FeTi−Cu composites revealed a Cu mass loss of 98.2 ± 0.9 %, indicating complete and highly selective dissolution of Cu, while the FeTi ligaments exhibit a significantly lower susceptibility to the etchant. Complete dissolution of the sacrificial Cu phase was confirmed. In the following, the materials obtained by etching from nc and ufg FeTi−Cu composites will be referred to as nanoporous (np) and ultra-fine porous (ufp) FeTi, respectively.

The mass loss during etching aligns with the SEM images in **Figure 3 (a-d)**, showing a more open and porous structure with prolonged immersion of the nc (np) FeTi−Cu composites. Analysis of the micrographs confirms the increased pore fraction, but shows only minor changes in ligament and pore sizes. The reasons for the more pronounced attack on the np compared to the ufp material could be (i) the higher specific surface area of the nc material and (ii) the more defective character and, in particular, higher curvature of the ligaments. The ligament sizes of the np and ufp materials are 32.8±0.4 nm and 338±3 nm, respectively. **Figure 3 (j)** also includes the FeTi grain sizes determined via EBSD (ufp) and TKD (np), matching the ligament sizes and, therefore, again confirming that the ligaments, in both np and ufp states, on average, consist of a single FeTi grain.

SEM-derived EDX results show Cu alloying for both np and ufp FeTi, with 3.7 ± 0.2 at.% and 1.2 ± 0.2 at.% at $r \approx 3$ mm, respectively. A slight radial dependence, with higher Cu contents at higher radii, indicates that the origin is very likely mechanical alloying during HPT. Fe levels are slightly reduced compared to Ti, consistent with the literature, which indicates that Cu preferentially substitutes for Fe in the FeTi structure.[25] Additionally, the EDX measurements indicate a stable composition, with no changes as a function of either immersion time or etching depth. As Cu is removed, the FeTi comes into contact with the highly oxidizing etchant and (mostly) passivates. Therefore, no leaching effect and concomitant compositional changes are expected, as confirmed experimentally. The associated EDX results are plotted in **Figures S19-S20 (Supporting information),** and **Table S3 (Supporting information)** summarizes the EDX results from both SEM and TEM.

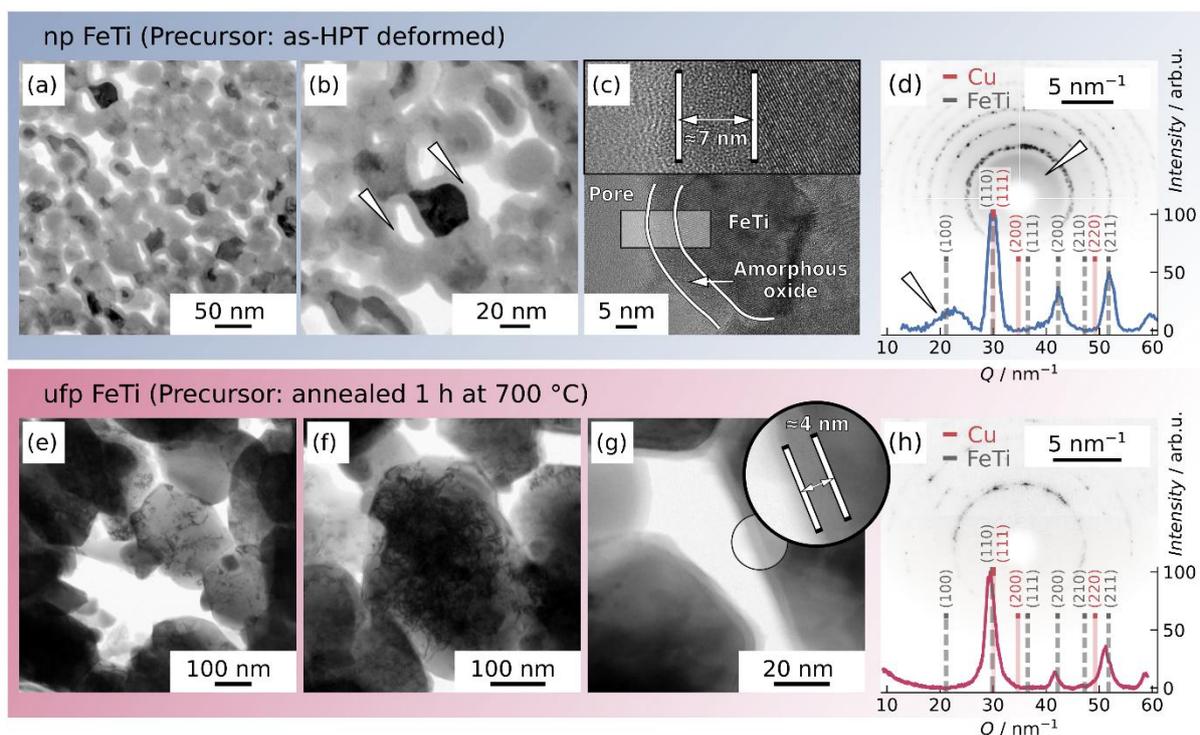

**Figure 4**: TEM micrographs and SAED patterns of (a-d) np FeTi and (e-h) ufp FeTi. (d,h) contain the raw and integrated SAED patterns; the scale bar shows the scattering vector $k$; the integrated data is plotted as a function of $Q = 2\pi k$. The arrows in (b) and (d) indicate the amorphous oxide layer in the TEM micrograph and the corresponding hump in the SAED diffractogram. The respective oxide thicknesses are marked in (c) and (g)

The TEM micrographs in **Figure 4** clearly show the presence of pores in both np and ufp FeTi. Close inspection also reveals the presence of an oxide layer surrounding the FeTi grains, and this core-shell structure of the ligaments is particularly pronounced in np FeTi. EDX maps in **Figures S21-S22** show elevated oxygen concentrations in these layers, corroborating their identification as oxides. These oxides have a relatively uniform thickness of 6.9 ± 0.5 nm and 3.9 ± 0.4 nm in np and ufp FeTi, respectively, in line with previous reports.[64,65] Despite this difference in thickness, the high-resolution (HR) TEM micrographs in **Figure S23** (**Supporting Information**) show that the oxide layer is amorphous in both the np and ufp FeTi.

The selected area electron diffraction (SAED) patterns in **Figure 4 (d,h)** contain rings associated with FeTi, but do not show signs of metallic Cu. In the SAED of np FeTi, highlighted by an arrow in **Figure 4 (d)**, an amorphous halo can be identified at lower $Q$ values, with the integrated pattern consequently exhibiting a broad hump at ≈ 22 nm$^{-1}$. This peak is associated with the amorphous oxide layer and is also visible in the SR-XRD data. A comparison of the SR-XRD, SAED, and fast Fourier transformed (FFT) HRTEM data in **Figure S24** shows a good agreement between the various methods. Due to the far lower specific surface area and, therefore, lower oxide volume fraction in ufp FeTi, the corresponding amorphous peak could not be resolved by SR-XRD and SAED. The formation of an amorphous oxide results from the strongly oxidizing conditions during etching, and this layer can be regarded as a barrier for hydrogen absorption. However, ufp FeTi also contains many dislocations, as seen, for example, in **Figure 4 (f)**, which might promote hydrogen absorption.

The pore sizes and morphologies of two selected np materials were investigated after 8 h and 24 h etching, respectively, using N$_2$ gas adsorption and desorption measurements at 77 K, with the details provided in **Figure S25 and Table S4** (**Supporting Information**). The obtained isotherms are of Type IV (a) with H1 hysteresis loops, irrespective of etch times, and indicate mesopores with a diameter > 4 nm and a cylindrical pore morphology.[66] The quantitative evaluation of the isotherms reveals a significant increase in pore volume, pore size, and specific surface area with prolonged etching, which aligns with the qualitative observations in the SEM micrographs in **Figure 3 (b-d)**. SEM seems to overestimate the size of the structural features, as the respective sizes are already approaching the limits of SEM resolution capabilities. However, **Figure S25 (c)** (**Supporting Information**) confirms that the global pore size distributions obtained from N$_2$ gas sorption match the pore sizes and Cu grain sizes determined via TEM and TKD.

As a first approximation, ligament sizes in np and ufp FeTi were estimated from the positions of the Bragg-like shoulders in the SANS/USANS scattering profiles, plotted in **Figure S26** (**Supporting Information**), using the relation $D ≈ Q/2π$. Np FeTi exhibits ligament dimensions in the range of 25 to 50 nm, whereas ufp FeTi shows substantially coarser features, with ligament sizes ranging from 200 to 500 nm, again confirming the SEM and TEM results.

**Functional characterization of the porous FeTi**

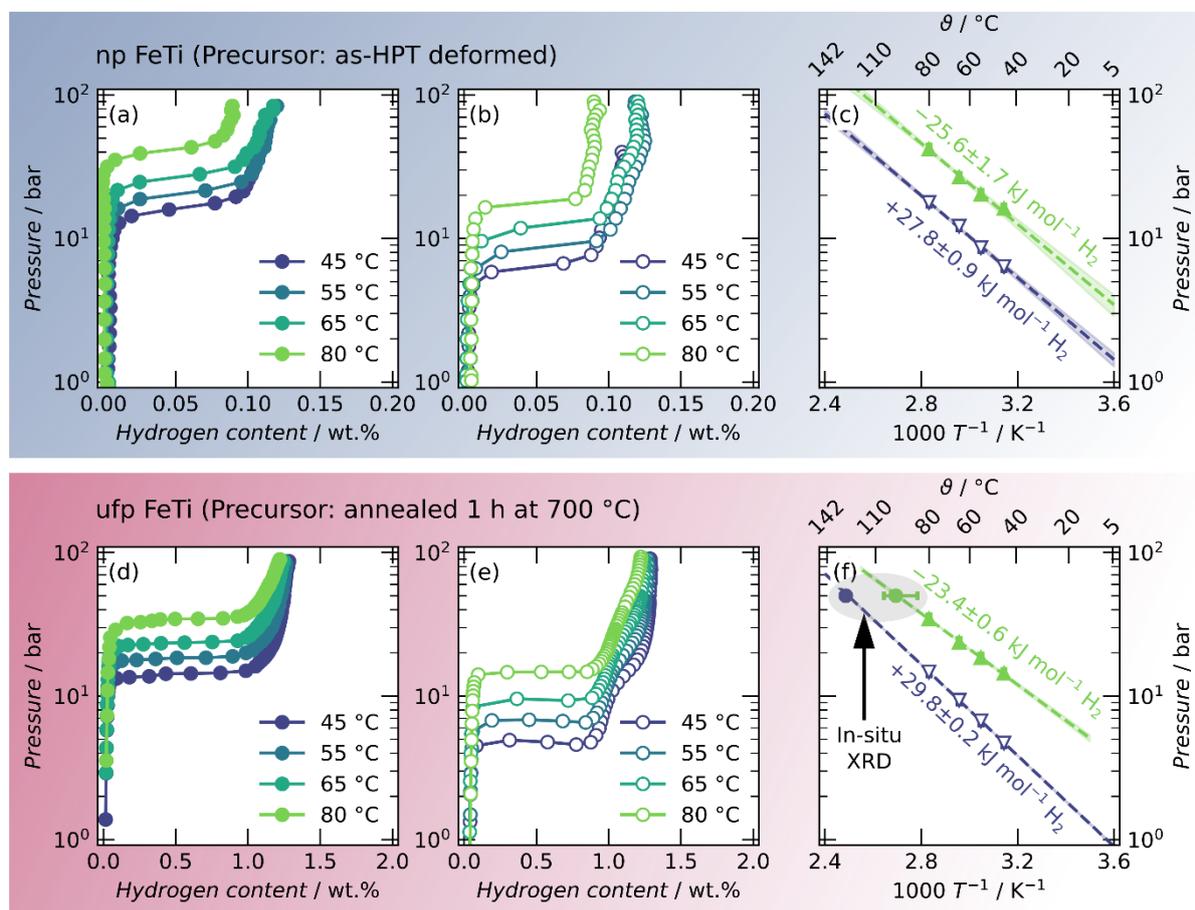

**Figure 5:** PCI measurements of (a-c) np and (d-f) ufp FeTi. (a,d) Absorption and (b,e) desorption branches of the PCI curves at 45 °C, 55 °C, 65 °C, and 80 °C, and the (c,f) derived Van't Hoff plots. Reported enthalpies and entropies were derived from the centers of the plateau regions, i.e., 0.05 wt.% and 0.5 wt.% hydrogen content, respectively. The arrow in (f) indicates values derived from absorption and desorption events observed using *in-situ* SR-XRD measurements.

Following the structural investigation, the hydrogen absorption and desorption properties were investigated by measuring PCI curves. The results, including the associated Van't Hoff plots, are presented in **Figure 5** for (a-c) np and (d-e) ufp FeTi, respectively. Despite the large surface area of np FeTi, the material does not absorb appreciable quantities of hydrogen at RT.

Nevertheless, following an activation treatment, as detailed in **Figure S27**, the subsequent PCI measurements, depicted in **Figure 5 (a,b)**, show well-defined absorption and desorption plateaus, but at a drastically reduced absorption capacity of about 0.12 wt.% at 45 °C and 90 bar. The pressure plateaus, albeit at reduced capacity, align well with the plateaus observed for conventional FeTi and were therefore related to the formation of the monohydride. Van't Hoff plots reveal that the enthalpies of hydride formation/decomposition, derived from the absorption/desorption branches, are $-25.6 \pm 1.7$ kJ mol$^{-1}$ H$_2$ and $+27.8 \pm 0.9$ kJ mol$^{-1}$ H$_2$, respectively. The associated entropy values are $-103.3 \pm 5.0$ J mol$^{-1}$ K$^{-1}$ H$_2$, and $+102.3 \pm 2.7$ J mol$^{-1}$ K$^{-1}$ H$_2$, respectively. These results are in good agreement with the values reported in literature, which are about ±24-30 kJ mol$^{-1}$ H$_2$ and ±106 J mol$^{-1}$ K$^{-1}$ H$_2$, for the α/β plateau, i.e., the formation/decomposition of the monohydride.[16,67–70]

Compared to np FeTi, the PCI curves of ufp FeTi in **Figure 5 (d-f)** show significant hydrogen absorption, with hydrogen contents of up to 1.29 wt.% at 45 °C and 90 bar. Notably, absorption occurred without a dedicated activation; the sample was only subjected to degassing at 100 °C under vacuum. Activation following the measurements depicted in **Figure 5** did not reveal significant changes in the PCI curves. The recorded isotherms exhibit well-defined pressure plateaus and indicate good reversibility. The Van't Hoff plot in **Figure 5 (f)** yields an enthalpy of hydride formation/decomposition of $-23.4 \pm 0.6$ kJ mol$^{-1}$ H$_2$ and $+29.8 \pm 0.2$ kJ mol$^{-1}$ H$_2$ as derived from the absorption/desorption branches, respectively. The entropy was determined as $-95.5 \pm 1.9$ J mol$^{-1}$ K$^{-1}$ H$_2$ and $+106.8 \pm 0.6$ J mol$^{-1}$ K$^{-1}$ H$_2$, respectively. Absorption and desorption conditions derived from the *in-situ* SR-XRD investigation, as detailed below, are in excellent agreement with the Van't Hoff plot, as indicated in **Figure 5 (f)**.

Following the PCI measurements, 10 absorption/desorption cycles were performed, confirming stable hydrogen storage properties, albeit with a slight decrease in capacity. The kinetic analysis showed fast kinetics with 50 % capacity reached after ≈ 15 s and agreed with the Johnson-Mehl-Avrami-Kolmogorov (JMAK) kinetics model for 1D growth.[46,71–73] For details, see **Figures S28-S29**.

The PCI curves of np and ufp FeTi exhibit both distinct differences and notable similarities. Both display a pressure plateau, and the determined enthalpies of hydride formation match the expected range for FeTi. The most striking difference is the drastically increased hydrogen uptake in ufp FeTi compared to the np material. Due to the low capacity of the latter, it is concluded that only a small fraction of the np material converts to the hydride, while the majority of the material remains inactive to hydrogen.

To better understand the hydrogen absorption behavior of porous FeTi and unambiguously confirm the postulated hydride formation, *in-situ* SR-XRD measurements were performed. The results associated with np and ufp FeTi are given in **Figures 6** and **7**, respectively.

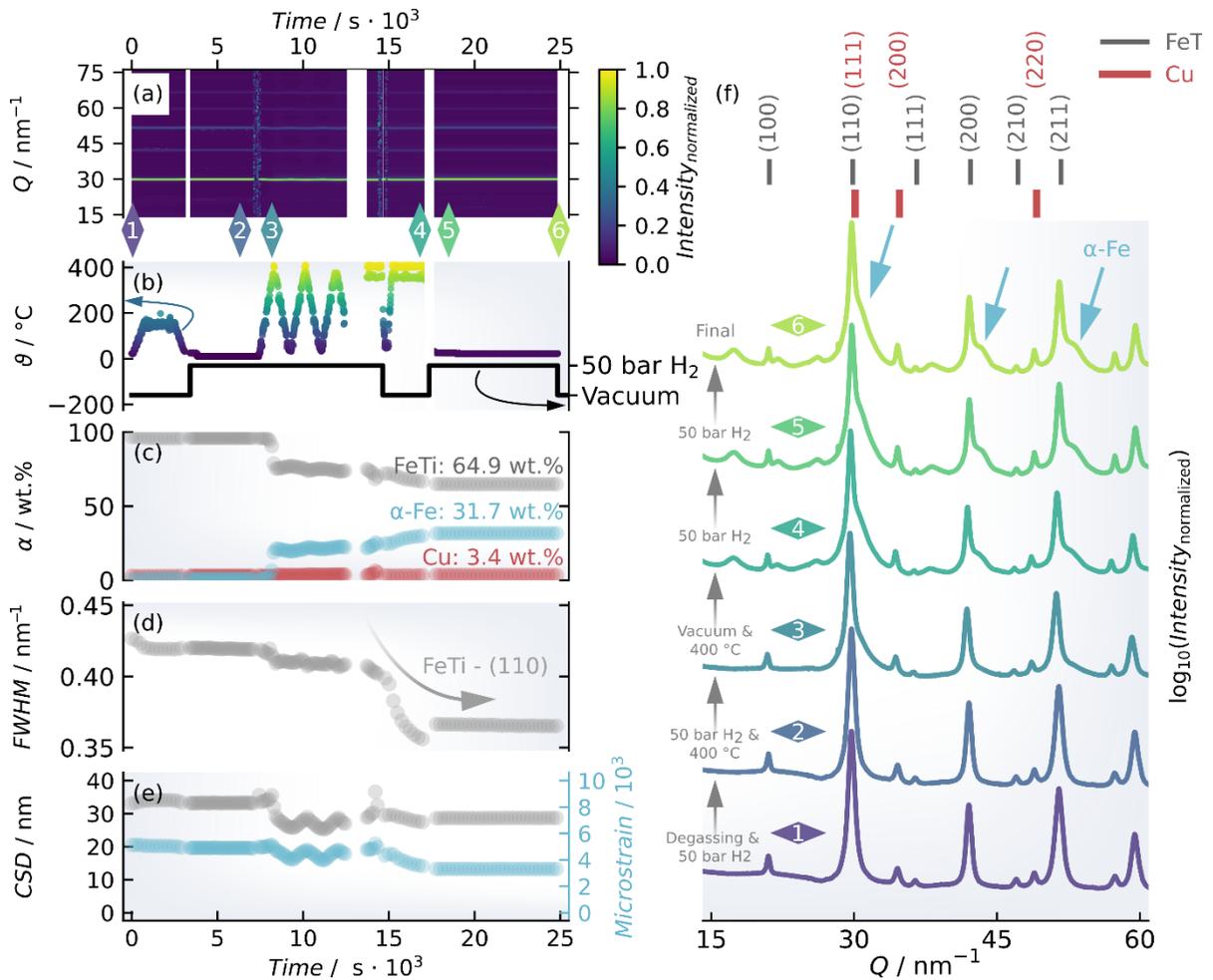

**Figure 6:** (a) 2D plot of the *in-situ* SR-XRD patterns obtained from np FeTi and (b) the associated hydrogen pressure and temperature conditions as a function of time. (c) Phase fractions α, (d) full width at half maximum (FWHM) of the FeTi (110) peak, and (e) the CSD sizes and microstrains. (f) SR-XRD patterns from distinct positions as indicated in (a). Note the logarithmic scaling in (f).

Initially, the SR-XRD patterns of np FeTi in **Figure 6** contain FeTi Bragg peaks along with weak reflections from the remaining Cu (3.4 wt.%). Subsequently, the material was subjected to 50 bar of hydrogen at RT, but did not form any hydrides. As seen in **Figure 6 (b)**, attempted activation by short-term heating to 400°C under vacuum and/or 50 bar hydrogen did not induce any hydride formation. However, when heated to 400 °C, the FeTi peaks developed shoulders, as indicated by arrows in **Figure 6 (f)**. These could be associated with the formation of (nc) α-Fe, a reaction product formed during the activation of FeTi.[23,74] Earlier works devised this as an important step during activation, induced by enhanced surface segregation at higher temperatures.[23] Other studies claimed it to be due to oxygen impurities in the hydrogen gas.[74,75]

Additionally, the broad, amorphous peak at low Q ranges disappears, and new, distinct peaks appear. These might be associated with one or more crystalline $FeTiO_x$, $TiO_x$, or $FeTiO_xH_y$ species, with the best agreement being with a $Ti_4Fe_2O_xH_y$ type compound.[76,77]

These changes induced in the oxide layers suggest that (surface) activation of FeTi did occur, yet this did not lead to hydride formation. Notably, as indicated by the reduction of the full width at half maximum (FWHM) during this measurement, the material relaxes during the attempted activation procedure, while hydrogen absorption should actually induce higher microstrains in the material.[78]

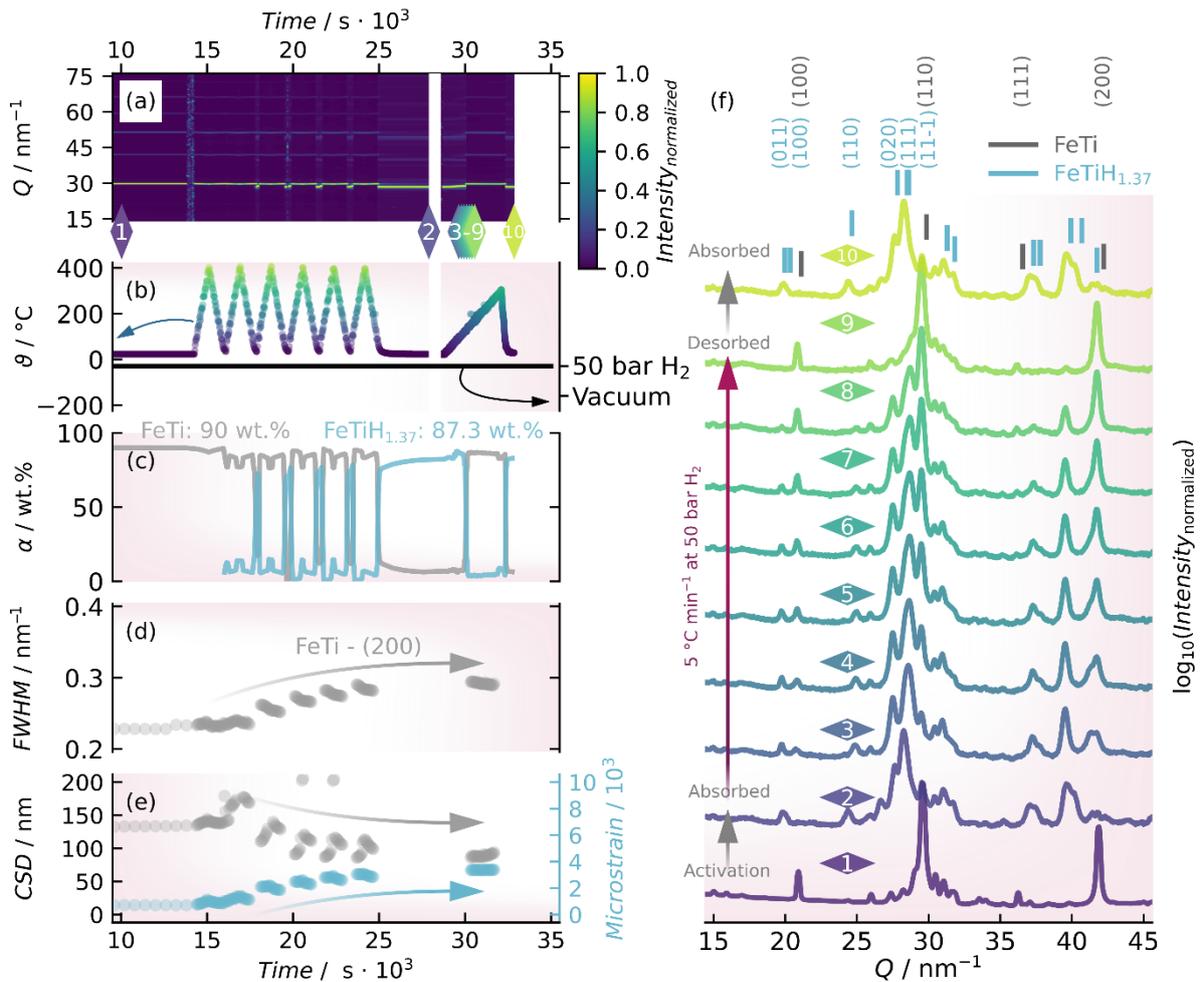

**Figure 7:** (a) 2D plot of the *in-situ* SR-XRD patterns obtained from ufp FeTi and (b) the associated hydrogen pressure and temperature conditions as a function of time. (c) Phase fractions α, (d) FWHM of the FeTi (200) peak, and (e) the CSD and microstrains. (f) SR-XRD patterns from distinct positions as indicated in (a). Note the logarithmic scaling in (f).

In contrast to the np material, ufp FeTi transformed to the FeTiH$_{1.3}$ monohydride after activation by short-term heating to 400°C at 50 bar hydrogen, with up to 87 wt.% of the material converting to the corresponding hydride. Temperature cycling resulted in reversible absorption/desorption at 129 ± 2°C/ 98 ± 10°C at 50 bar. These temperatures are included in the Van't Hoff plot in **Figure 5 (f)**, matching the result of the PCI measurements. **Figure 7 (f)** presents selected SR-XRD patterns as indicated in **Figure 7 (a)**. In particular, the intermediate patterns show the desorption induced by heating.

As seen in **Figure 7 (d)**, multiple temperature-induced absorption/desorption cycles resulted in an increased FWHM of the FeTi peaks in the desorbed state, indicating that microstrains built up during cycling. This is also reflected in the Rietveld refinement results in **Figure 7 (e)**, with both slightly reduced CSD sizes and increased microstrains. Such a build-up of microstrains and grain refinement is consistent with the well-known phenomenon of disproportionation, i.e., self-pulverization, during cycling. However, although the bulk materials fractured significantly during recording the PCI curves, the resulting particles maintained their underlying ufp structure, as seen in **Figure S30 (Supporting Information)**. **Figure S30 (f)** indeed shows that the ligament sizes remain stable upon hydrogen loading and unloading, while the internal crystalline size, as determined from EBSD, refines slightly. Altogether, this indicates a certain degree of microstructural and mechanical stability of ufp FeTi, even over multiple absorption/desorption cycles.

# Discussion

Combining the PCI curves with the *in-situ* SR-XRD results yields two main conclusions: (i) np FeTi forms only a limited amount of hydride phase, despite subtle changes that indicate activation. (ii) Despite identical etching and a similar oxide layer, ufp FeTi does absorb hydrogen, forming the corresponding monohydride and exhibiting stable and repeatable absorption and desorption behavior.

**Negligible hydride formation in nanoporous FeTi**

As indicated by the PCI curves in **Figure 5,** a small fraction of np FeTi converts to the respective hydride. The related enthalpy of hydride formation provides a clue in this regard. Specifically, the value agrees with plain FeTi, but not with Cu-substituted FeTi, which exhibits a higher (absolute) enthalpy of hydride formation.[25,26,79–81] Cu tends to substitute Fe in the ordered B2 structure, expanding the lattice and lowering the enthalpy of hydride formation. Values of $-37.2$ kJ mol$^{-1}$ H$_2$ are reported for up to 10 at.% Cu; (Mn, Cu)-substituted FeTi yields enthalpies as low as $-37.6$ kJ mol$^{-1}$ H$_2$. EDX data collected both in SEM and TEM indicate a Cu content of up to 4 at.% in np FeTi. Extrapolating the values from Shinar *et al.*[81] to a Cu content of 4 at.% yields an enthalpy of about $|31|$ kJ mol$^{-1}$ H$_2$, significantly higher than the values of np FeTi, namely $-25.6$ kJ mol$^{-1}$ H$_2$ and $+27.8$ kJ mol$^{-1}$ H$_2$. However, revisiting **Figures 1 and S6** reveals that relatively large FeTi particles, neither fragmented during HPT nor subjected to appreciable mechanical alloying, exhibit a negligible Cu content. This is particularly visible in the associated EDX maps in **Figures S15**-**S16 (Supporting information)**.

Therefore, the low capacity, as well as the subtle changes in thermodynamics, can be explained by assuming that only the larger FeTi particles formed hydrides, while nanoscale ligaments, containing about 4 at.% Cu, do not. Still, Cu-substitution does not hinder but rather favors hydride formation. The reason for this impairment must thus be found in the material's microstructure and morphology, not its composition.

**FeTi oxide – Barrier for hydrogen diffusion**

It is well established in literature that the primary obstacle to hydrogen absorption in FeTi is the surface oxide layer, which can be overcome only by an activation treatment at high temperatures and hydrogen pressures. Due to the nature of the etching process, the ligaments in both np and ufp FeTi are covered by an oxide layer with 7 nm and 4 nm thickness, respectively. The question is whether this oxide layer poses a barrier to hydrogen diffusion or if other processes are responsible for the blocked hydride formation. Since the ufp FeTi did absorb hydrogen without activation despite the presence of an oxide layer of similar structure and morphology, it is unlikely that it acts as an insurmountable diffusion barrier for hydrogen. Especially in the case of np FeTi, it is rather likely that the oxide should be relatively permeable for hydrogen due to the high surface area, curvature, defect and interface density. Additionally, *in-situ* SR-XRD of np FeTi confirmed subtle changes in the oxide layer, indicating activation for hydrogen absorption.

The surface permeability was indirectly confirmed through the use of electrochemical loading of np and upf FeTi, a comparable aggressive method, with details provided in the supporting information. XRD patterns, given in **Figure S31 (Supporting Information),** recorded immediately following electrochemical hydrogen loading, show the formation of hydrides for both np and ufp FeTi. Np FeTi showed fast desorption and decomposition of the hydride, while within ufp FeTi the hydride remained stable for extended periods (i.e., days), as expected for "oxide-stabilized" FeTi.[16] This suggests a higher permeability of the oxide layer for np FeTi, compared to the ufp material, and consequently, the oxide layer of np FeTi is not the limiting component to hydrogen absorption.

**Confinement effect**

Based on these results, we conclude that the reason for the blocked hydride formation of the nanoporous material is its distinct morphology, particularly the core-shell structure of the ligaments. As seen in the TEM micrographs in **Figure 4**, this core-shell structure consists of a hydrogen-absorbing FeTi core and a non-absorbing but permeable amorphous oxide shell. Notably, one aspect that has not been discussed so far is the necessity of accommodating the volume expansion associated with hydride formation and the associated stresses.

It is well known that stresses influence the behavior of hydrogen and its absorption in materials; for example, long-range attractive interactions between hydrogen atoms are mediated by the distortion of the host lattice.[82] Consequently, hydrogen cannot be considered independently of the accompanying elastic stresses, which are themselves strongly influenced by the microstructure.[83] For example, recent ab initio calculations also showed the impact of mechanical strains on the nucleation behavior of FeTi hydrides,[84] aligning with experimental orientation relationships between the nucleated hydride and the respective host metal.[85,86]

As FeTi expands during hydrogen absorption, the rigid oxide shell covering the ligaments in the np material poses a barrier to this expansion. As indicated by the evolution of the amorphous oxide phase during *in-situ* SR-XRD in **Figure 6**, activation can alter the internal structure and phase composition of the oxide layer, but it does not remove the covering layer as a whole. Therefore, we hypothesize that the presence of such a surface layer causes a *confinement effect* that severely blocks hydride formation. In fact, hydride formation could occur in larger FeTi ligaments and particles, as they have a more favorable core-to-shell size ratio. This aligns with the interpretation of the PCI curves of np FeTi, namely that only larger FeTi particles in the microstructure hydrogenate, while nanoscale ligaments do not. The ufp FeTi overall has a far more favorable core-to-shell size ratio, which relieves the confinement effect and allows hydride formation. This mechanism can explain the differences in sorption behavior of np and ufp FeTi, and is illustrated in **Figure 8 (a,b)**.

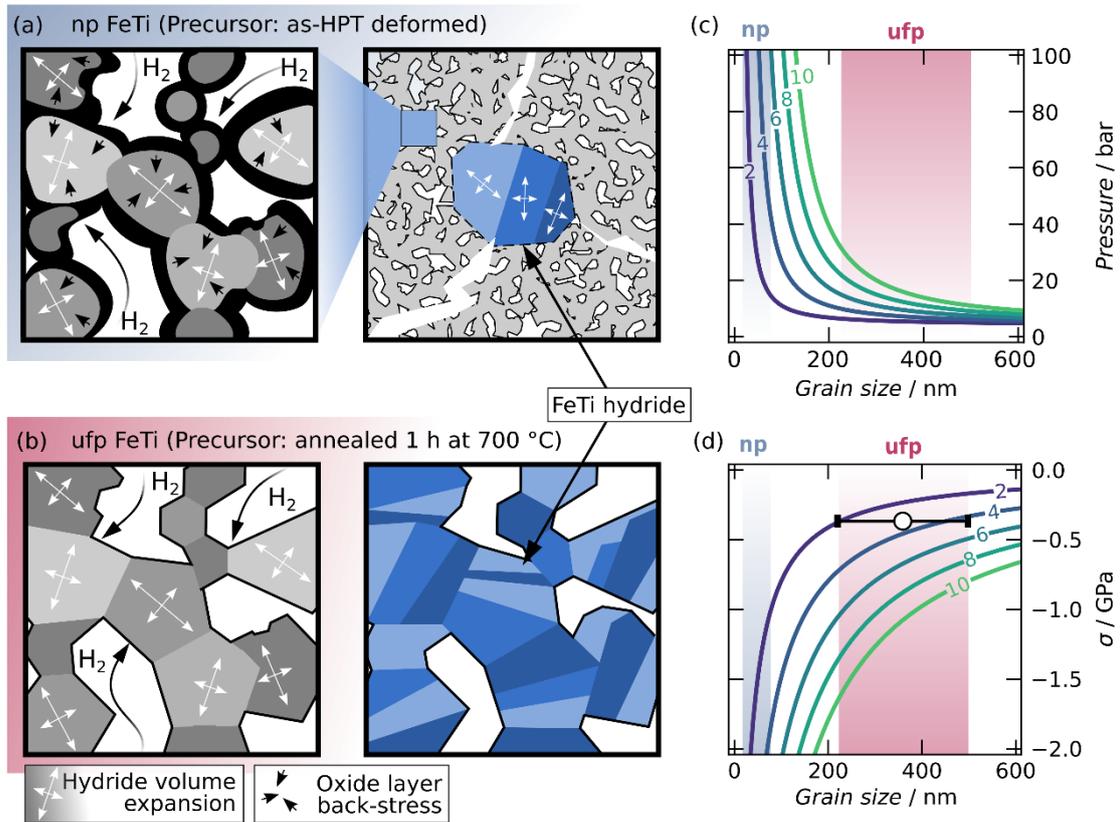

**Figure 8**: Illustration of the (a) blocked hydride formation in np and (b) hydrogen absorption in ufp FeTi. (c) Equilibrium hydrogen pressures and (d) mechanical backpressures calculated for the FeTi-oxide core-shell structures using an analytical model. (d) includes the stresses calculated based on the enlarged hysteresis in ufp FeTi. The respective equations are outlined in the Supporting Information. The numbers in (c,d) are the respective oxide layer thicknesses in nm.

In addition to these qualitative arguments, an analytical model was used to underscore the proposed confinement effect. Based on the works of Wagner and Horner,[87] Tessier *et al.* derived the elastic contribution to the chemical potential of the absorbed hydrogen in FeTi, exploring the impact of amorphous grain boundaries on the absorption behavior.[88] This analytical solution was slightly modified and applied to the current ligament structure, replacing the original absorbing amorphous grain boundary with the non-absorbing oxide layer. Knowing the volume expansion per absorbed hydrogen atom in the FeTi lattice ($\approx 2.8$ Å$^3$)[68,82,88], the relevant mechanical constants, and the core/shell sizes, this model yields a concentration-dependent equation for the hydrostatic backpressure on the absorbing FeTi core. This pressure markedly influences the chemical potential of hydrogen at interstitial sites and, consequently, the related absorption pressure. The model details and equations are provided in the supporting information.

**Figure 8 (c)** shows the model results, which indicate a core-to-shell size ratio below which hydride formation becomes energetically strongly impeded. The rapid rise at lower ligament sizes originates from the exponential relation between chemical potential and absorption pressure. Specifically, np FeTi is situated in a size regime where hydride formation becomes highly unfavorable, explaining the negligible hydrogen uptake of the np material. In contrast, ufp FeTi occupies a regime in which the core–shell morphology has only a small destabilizing effect on hydride stability. Any slight destabilization is, however, masked by concurrent Cu substitution, resulting in a net hydride stabilization for ufp FeTi.

However, the observed absorption-desorption hysteresis of ufp FeTi could be used to additionally validate the model results. Specifically, according to the Schwarz–Khachaturyan model, the hysteresis arises from elastic stresses generated by forming a coherent hydride phase.[89,90] In np and ufp FeTi, these intrinsic stresses are further modified by the backpressure from the core–shell morphology. Under the assumption that, in the ufp material, intrinsic (coherent-domain) and extrinsic (core–shell) stresses add linearly, the elastic stresses can be estimated based on the experimentally measured hysteresis. Subsequently, these results can again be compared with the experimentally observed structure and the model results. Using the recorded PCI curves in **Figure 5**, the hysteresis-free energy $\Delta G_{hyst}$ was calculated using

$$\Delta G_{hyst} = RT \ln\left(\frac{p_{abs}}{p_{des}}\right) \qquad . \qquad (3)$$

Subtracting the intrinsic hysteresis free energy of FeTi[91] yields an excess energy originating from the backpressure due to the core-shell structure. Using this excess free energy and performing the model calculations in reverse, i.e., from energy to stress, yields the excess elastic stress within the FeTi grain during hydride formation, amounting to approximately 358±14 MPa. Plotting this stress together with the median ligament size in **Figure 8 (d)** yields a remarkable agreement with the results of the analytical model.

Consequently, the discussed results show that the elastic stresses induced by the core-shell structure of porous FeTi significantly influence the hydride formation as well as the associated hysteresis. Indeed, Kobayashi *et al*. reported minor hydrogen absorption in FeTi nanopowders, which could have experienced a similar blocking confinement effect.[47,52] Such confinement has been used deliberately to destabilize Mg hydrides,[92–95] and recently published mesoscale modeling confirms the destabilization of $MgH_{2x}$ embedded in a host material.[96] Indeed, the model used in the simulations resembles the core-shell structure observed in this study, highlighting the universal impact of such morphology on metal hydrides.

This study, therefore, highlights the importance of a system's micromechanical response on metal hydride nucleation and growth. It particularly demonstrates that stresses, governed by the intrinsic microstructure and morphology of porous materials, significantly impact the hydrogen absorption in np and ufp FeTi. It is key that such influences are considered for materials with similar structural features.

# Conclusions

In this study, we prepared bulk porous FeTi by selectively dissolving a sacrificial Cu phase from a FeTi−Cu composite material obtained by severe plastic deformation using high-pressure torsion. By applying heat treatments, the grain sizes of the composite can be varied in a controlled manner. The complete dissolution of metallic Cu from the composites with set grain sizes allows the preparation of nanoporous and ultra-fine porous FeTi, with well-defined ligament sizes on the order of 50 and 300 nm, respectively. However, due to the nature of the etch process, the FeTi ligaments are covered by a 4-7 nm thick amorphous oxide layer.

The hydrogen sorption properties were probed by PCI measurements and *in-situ* SR-XRD investigations. These experiments revealed that np FeTi neither absorbs appreciable amounts of hydrogen nor forms large amounts of hydride, despite the large surface area of the np material and repeated activation treatments. Conversely, ufp FeTi with larger ligaments absorbs hydrogen well and forms the respective monohydride $FeTiH_{1.3}$.

The results can be rationalized by considering the effect of the oxide layer covering the FeTi ligaments, constituting a core-shell structure with an absorbing FeTi core and a non-absorbing oxide shell. The backpressure induced by hydrogen absorption in this core-shell morphology blocks any hydride formation for unfavorable, i.e., low, core-to-shell size ratios. In ufp FeTi, where larger grains and thinner oxide layers are present, hydride formation is not blocked; however, the observed slightly increased hysteresis in PCI measurements can be rationalized by a not insignificant backpressure of 358 MPa.

These insights bear important implications for the design of any nanostructured hydrogen storage material with respect to the desire for fine nanoporous structures to enable short diffusion pathways, balanced by the need to maintain phase transformation in a potentially oxygen-containing environment.

# Acknowledgments

We acknowledge DESY (Hamburg, Germany), a member of the Helmholtz Association HGF, for the provision of experimental facilities. Parts of this research were carried out at PETRA III using beamline P02.1. Beamtime was allocated for proposal I-20230216 EC. We acknowledge the support from Felix Römer and Eray Yüce. The authors also thank Silke Kaufmann for her efforts in preparing the TEM samples. This research activity is part of the Strategic Core Research Area SCoRe A+ Hydrogen and Carbon and has received funding from Montanuniversität Leoben. This work received funding from the Climate and Energy Fund Austria within the "Energieforschungsprogramm 2023", project "HEAfine4H2" [administered by FFG, grant number 914968]. The authors further gratefully acknowledge the financial support from the Austrian Science Fund (FWF) [P 34840-N]. We also greatly acknowledge the funding from the Dtec project – Digitalization and Technology Research Center of Bundeswehr. We also sincerely thank Dr. Jitendra Mata at the Kookaburra USANS instrument and Dr. Joshua King at the Quokka SANS instrument, Australian Nuclear Science and Technology Organisation (ANSTO), Lucas Heights, for their expert guidance and invaluable support throughout the neutron scattering measurements.

# Supporting Information

## FeTi−Cu composites: Structural characterization

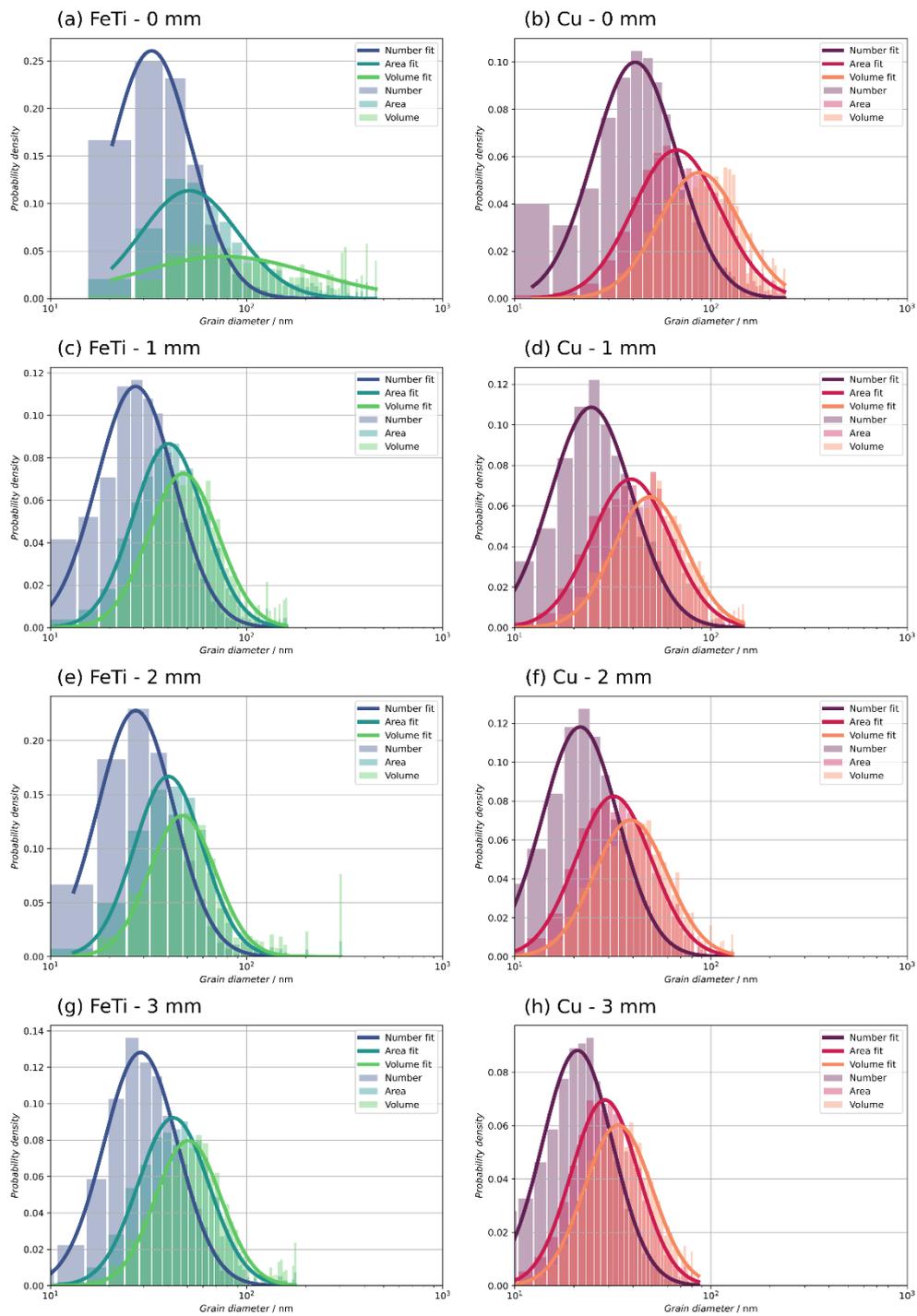

**Figure S1**: Grain size distributions of nanocrystalline (nc) FeTi−Cu composites as determined by SEM image analysis and weighted by grain number, area, and volume. (a, c, e, g) show the results for FeTi and (b, d, f, h) for Cu.

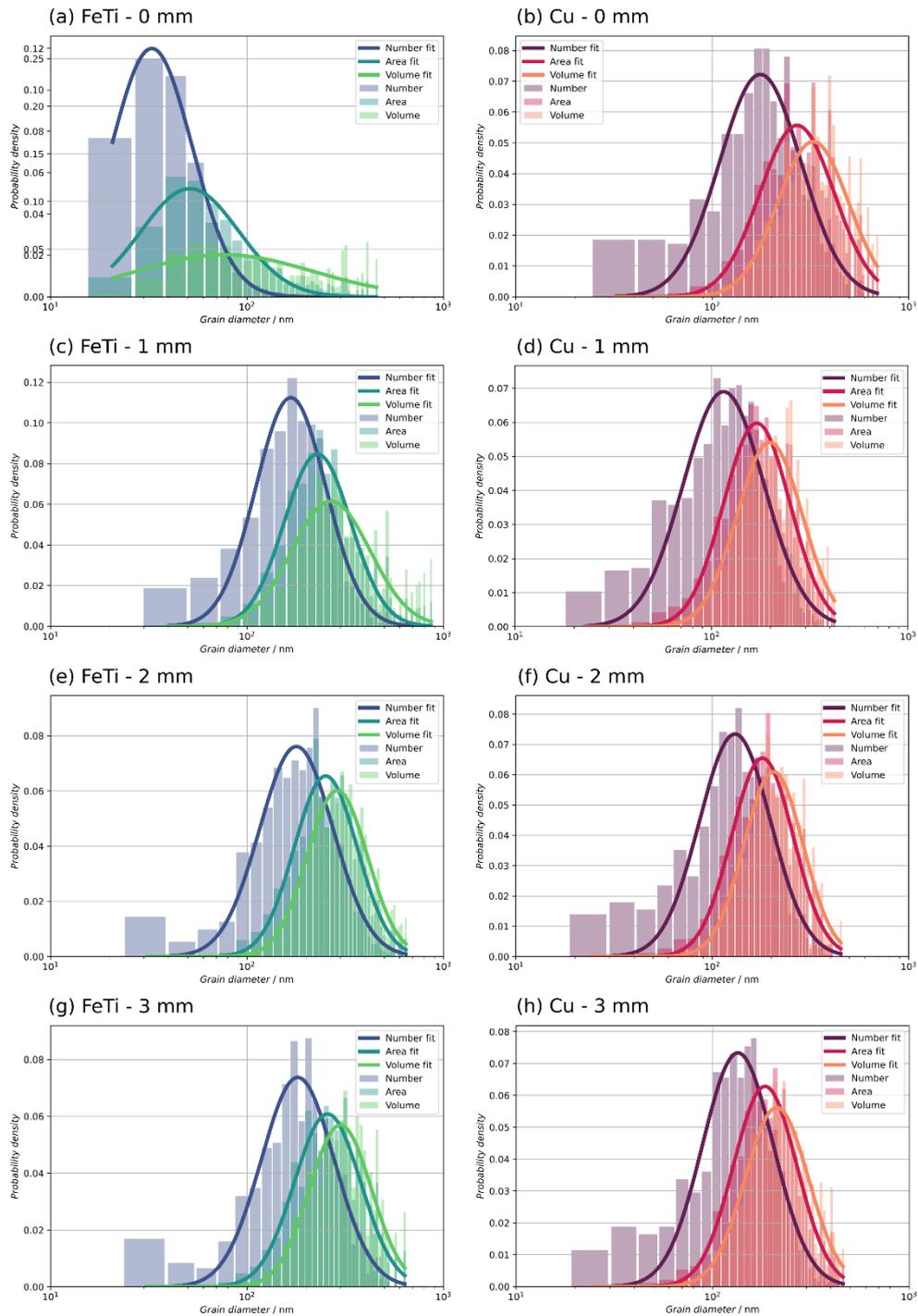

**Figure S2:** Grain size distributions of ufg FeTi−Cu composites, after annealing nc FeTi−Cu composites for 1 h at 700 °C, as determined by SEM image analysis and weighted by grain number, area, and volume. (a, c, e, g) show the results for FeTi and (b, d, f, h) for Cu.

**Table S1:** FeTi and Cu grain sizes of nc and ufg FeTi−Cu composites. The mean grain size was determined by fitting a log-normal distribution. Area-weighted grain sizes are reported. All values are reported for $r \approx 3$ mm ($\gamma \approx 1900$). TEM values are not phase resolved.

| Phase | State | SEM | EBSD | TEM | TKD |
|---|---|---|---|---|---|
| FeTi | As-HPT | 54.8 ± 0.7 | - | 20.2 ± 0.2 | 33.7 ± 0.8 |
| Cu | As-HPT | 36.3 ± 0.2 | - | | 28.7 ± 1.0 |
| FeTi | Annealed | 322 ± 8 | 289 ± 12 | 239 ± 2 | 226 ± 8 |
| Cu | Annealed | 227 ± 3 | 178 ± 7 | | 104 ± 4 |

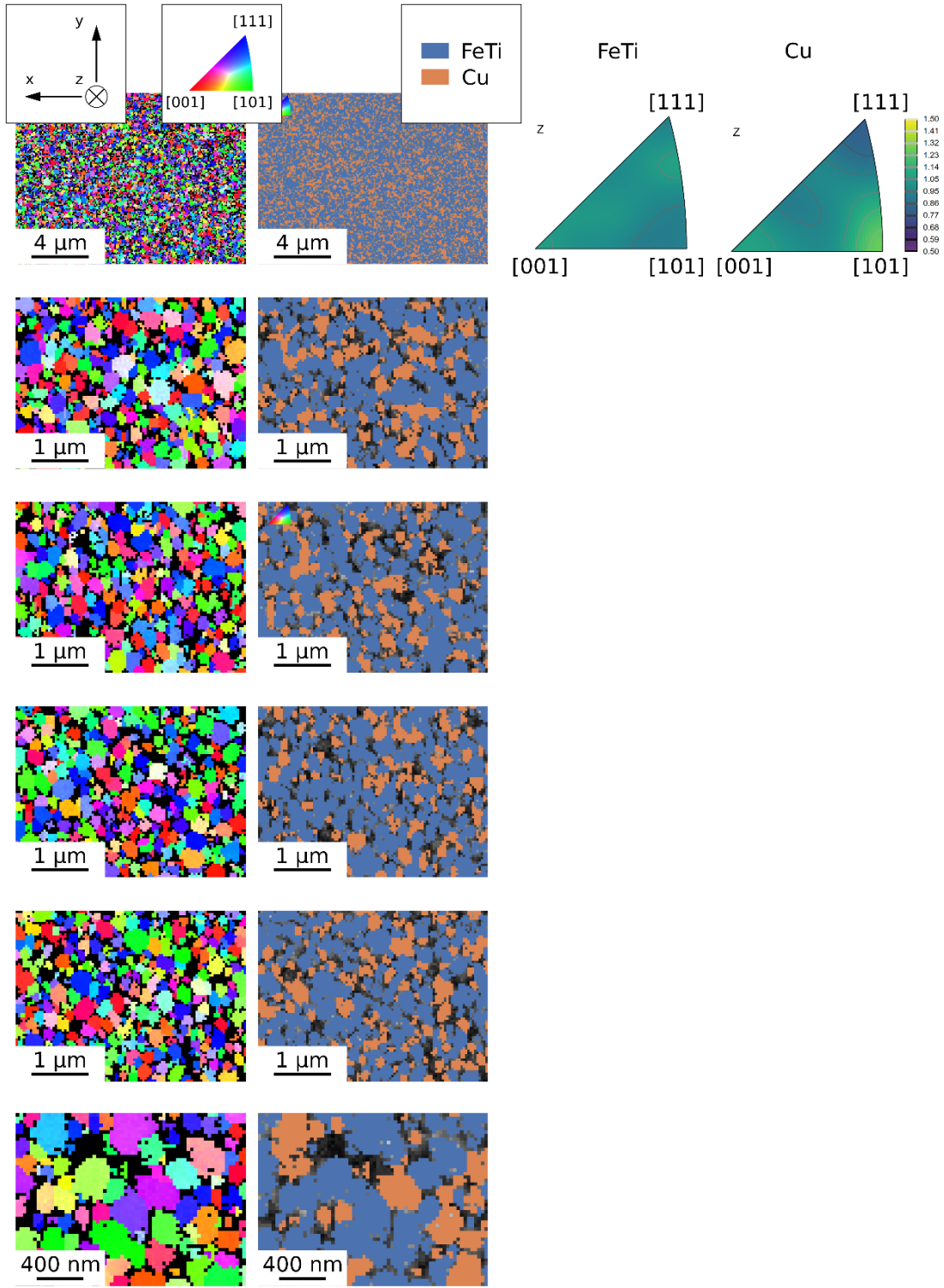

**Figure S3**: IPF maps, phase maps, and IPFs obtained by EBSD of ufg FeTi−Cu composites. All images were taken in tangential view at $r \approx 3$ mm ($\gamma \approx 1900$).

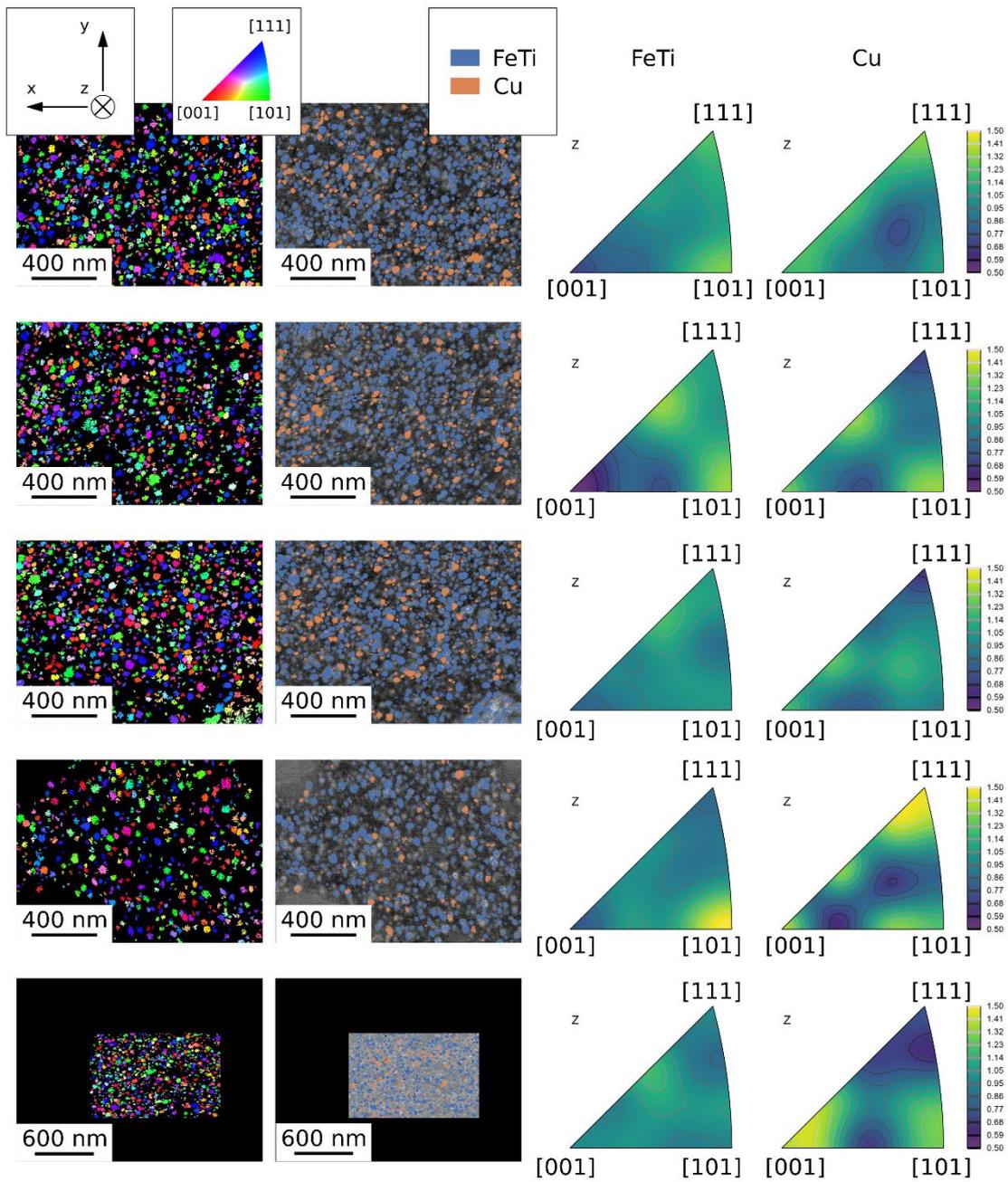

**Figure S4**: IPF maps, phase maps, and IPFs obtained by TKD of nc FeTi−Cu composites. All images were taken in axial view at $r \approx 3$ mm ($\gamma \approx 1900$).

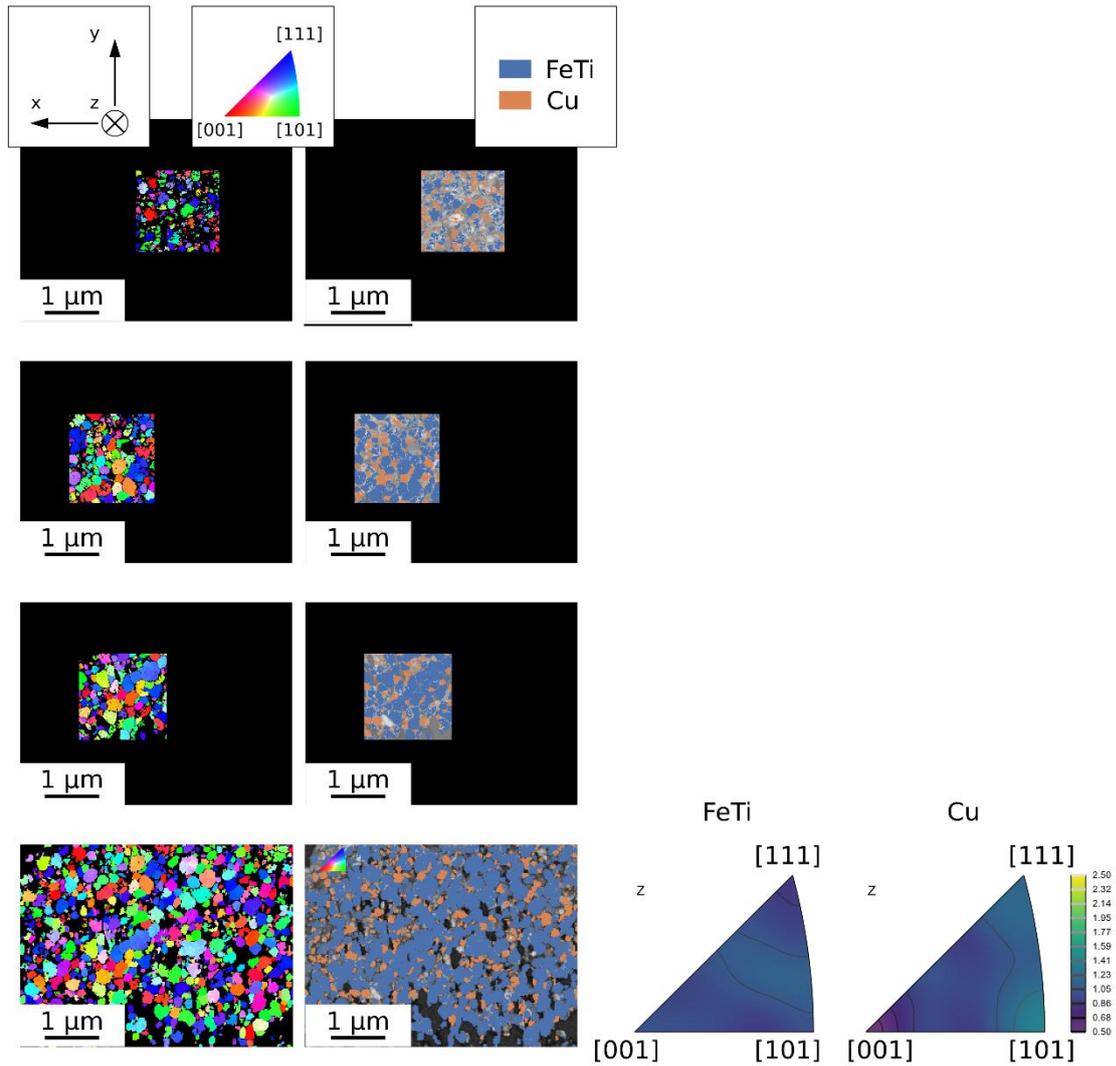

**Figure S5**: IPF maps, phase maps, and IPFs obtained by TKD of ufg FeTi−Cu composites. All images were taken in axial view at $r \approx 3$ mm ($\gamma \approx 1900$).

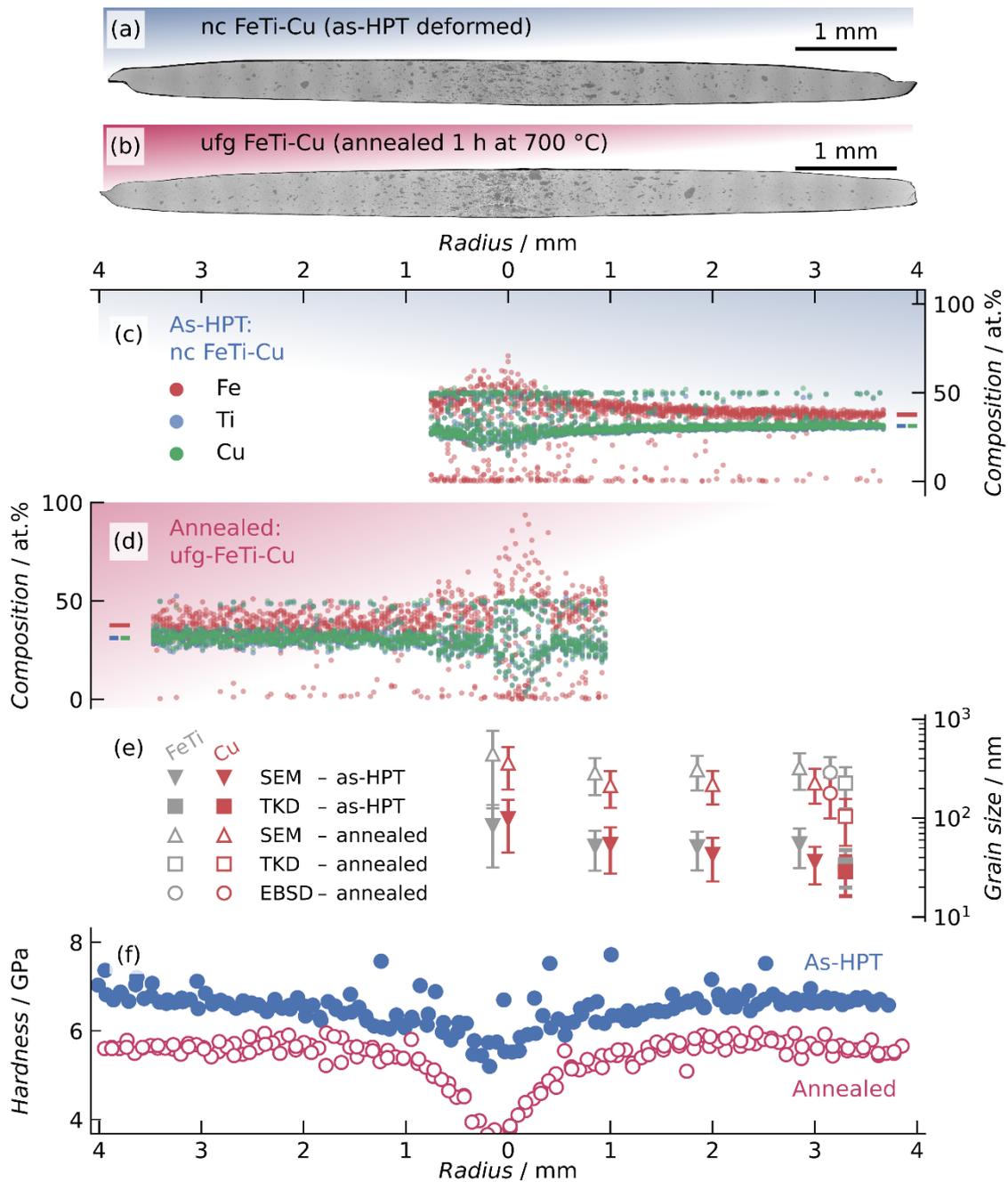

**Figure S6:** (a,b) BSE SEM micrographs of the cross-sections of nc and ufg FeTi−Cu composites providing an overview of the samples and (c,d) the radially resolved compositions probed via EDX. (e) Grain sizes determined via SEM, EBSD, and TKD. Note that the individual data points at $r = 3$ mm were shifted slightly along the x-axis to allow for easier readability. (f) Microhardness measurements along the radius of the HPT disks.

Overview SEM micrographs of the cross-sections of the HPT disks in **Figure S6 (a,b)** confirm a homogeneous microstructure at *r*>1 mm. Nevertheless, a few larger FeTi particles, insufficiently fragmented during deformation, are unavoidable in such material pairings.[1,2]

The compositions of the FeTi−Cu composites were determined via EDX along the radial direction and are given in **Figure S6 (c)**. At radii exceeding 1 mm, the values converge to the nominal global composition of the composite. In the regions close to the center, i.e., at r ≈ 0 mm, the EDX results scatter significantly, indicating a less refined and coarser microstructure. Nevertheless, outliers with a Cu content close to 0 at.% are still present at higher radii and are associated with the aforementioned larger FeTi particles. Overall, the Fe:Ti ratio is close to 1:1, as expected. Additional EDX maps and associated elemental compositions are provided in **Figures S7-S8** and **Table S2**.

The grain sizes determined by image analysis (phase identification followed by watershedding) of the SEM micrographs are provided in **Figure S6 (e)**. Values obtained by EBSD and TKD at *r* ≈ 3 mm align with these results. Generally, the grain size is stable, increasing notably only close to the disk center. **Figure 6 (f)** shows the microhardness obtained from the cross-sections, confirming the saturation of the mechanical properties at higher radii. At radii > 1 mm, nc FeTi−Cu composites have a hardness of 6.7 ± 0.4 GPa, while for ufg FeTi−Cu nanocomposites, the hardness reduces to 5.6 ± 0.2 GPa. This reduction is expected due to grain growth, resulting in a reduced Hall-Petch strengthening, as well as a general decrease in defect densities. The disk center exhibits a slightly lower hardness, but along most of the cross-section, the same hardness was observed, indicating the saturation of the mechanical properties and, therefore, associated microstructure. The chemical (Figure 6 (c,d)), structural (Figure 6 (e)), and mechanical (Figure 6 (f)) characterization of the composite suggests a saturated microstructure at r ≥ 1mm, about 94 % of the sample volume.

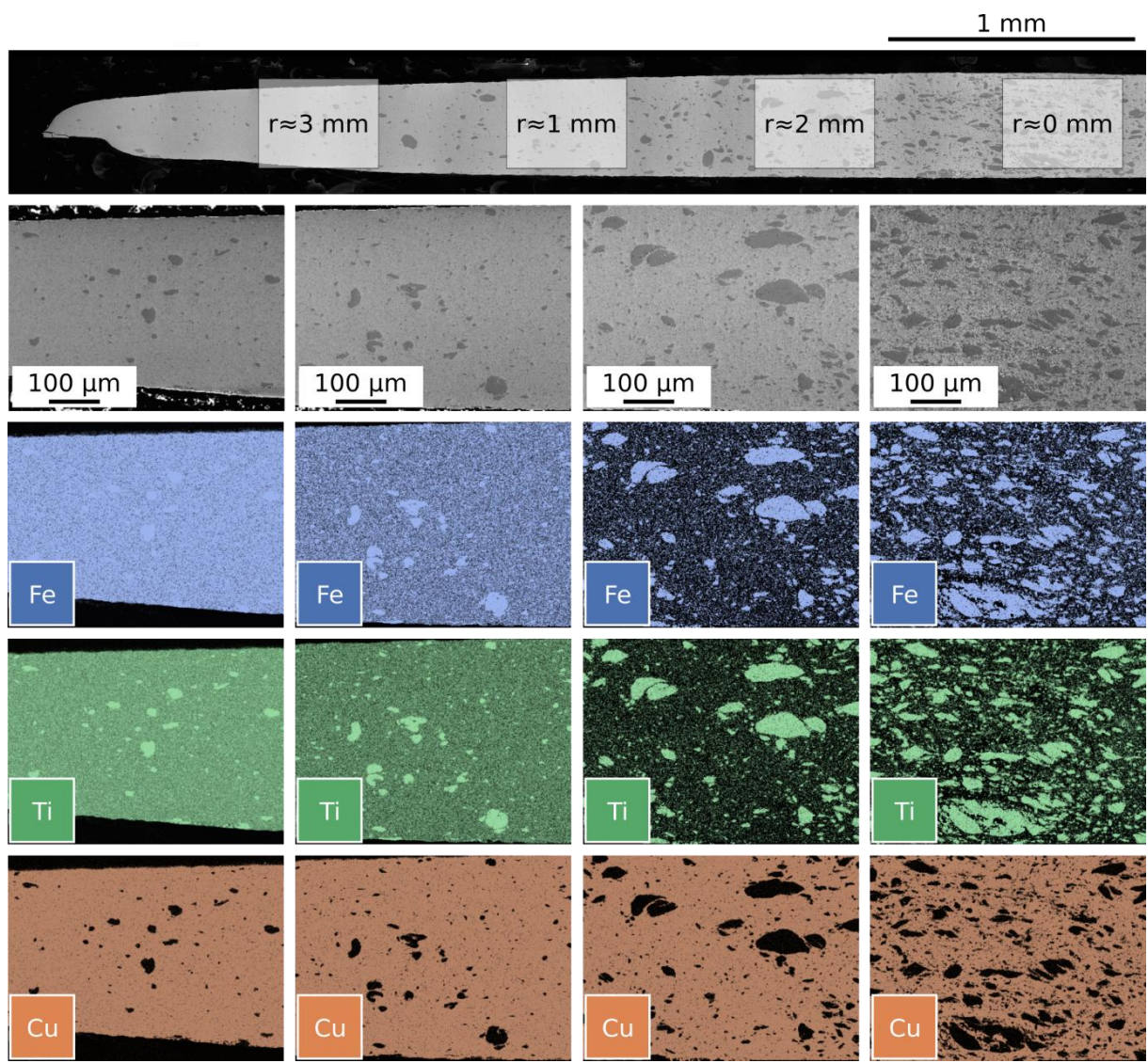

**Figure S7:** SEM EDX maps of nc FeTi−Cu. The respective global elemental compositions determined by quantifying the maps are given in **Table S2**.

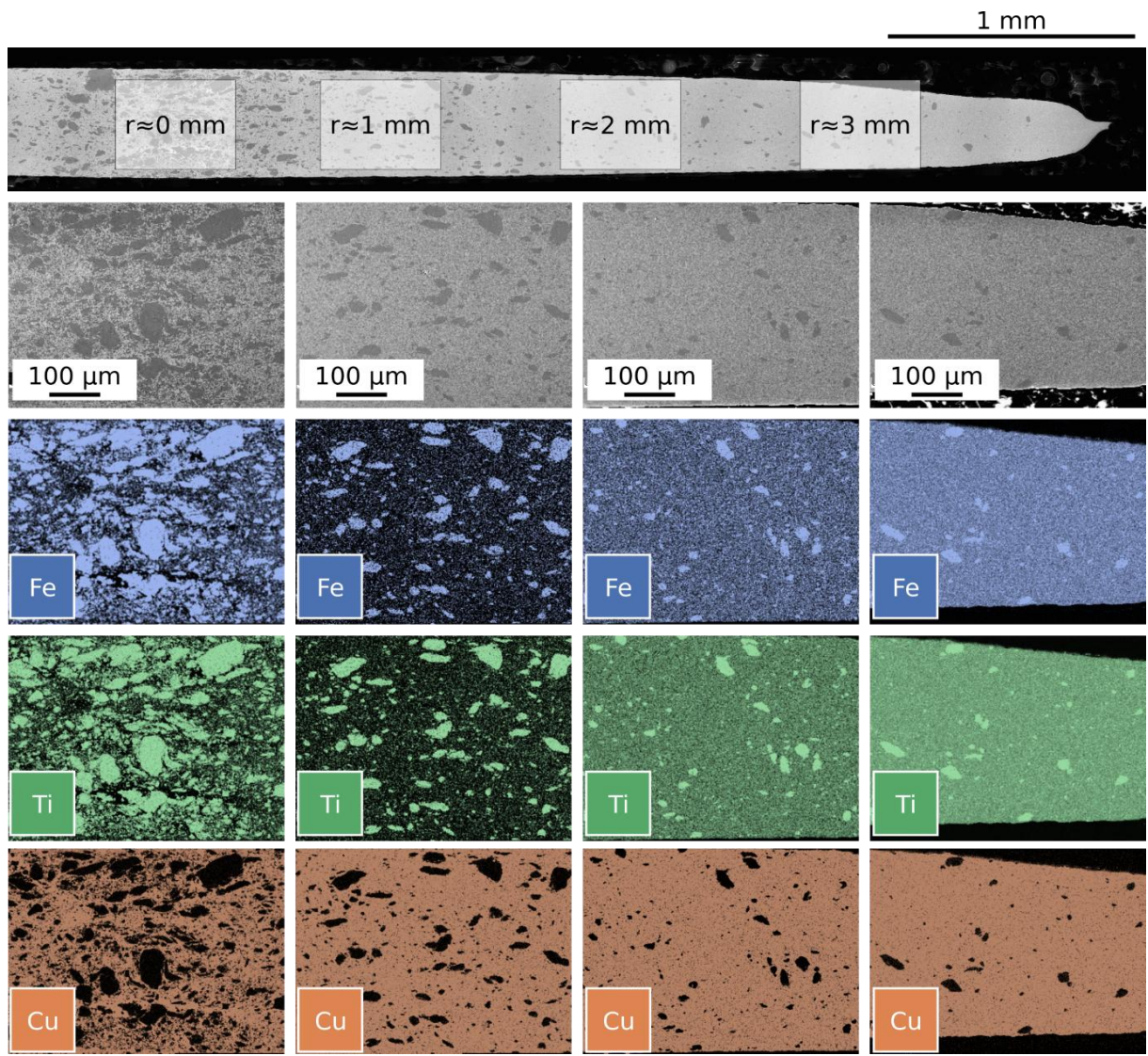

**Figure S8:** SEM EDX maps of ufg FeTi−Cu composites. The respective global elemental compositions determined by quantifying the maps are given in **Table S2**.

**Table S2**: Elemental composition of nc and ufg FeTi−Cu composites determined via SEM EDX. The compositional values were derived from the EDX maps in **Figures S7 and S8**.

| $r$ / mm | *Fe content* / at.% | *Ti content* / at.% | *Cu content* / at.% |
| --- | --- | --- | --- |
| As-HPT | | | |
| 0 | 32.7 ± 0.8 | 33.6 ± 0.7 | 33.7± 1.0 |
| 1 | 31.8 ± 0.8 | 32.4 ± 0.7 | 35.8 ± 1.0 |
| 2 | 31.2± 0.8 | 31.7± 0.7 | 37.1 ± 1.1 |
| 3 | 31.4± 0.8 | 31.9± 0.7 | 36.7 ± 1.1 |
| Annealed – 1 h at 700 °C | | | |
| 0 | 33.5 ± 0.9 | 34.7 ± 0.8 | 31.8 ± 0.9 |
| 1 | 32.2 ± 0.8 | 32.9 ± 0.7 | 35.0 ± 1.0 |
| 2 | 31.8± 0.8 | 32.4 ± 0.7 | 35.8 ± 1.0 |
| 3 | 31.9± 0.8 | 32.3± 0.7 | 35.7 ± 1.0 |

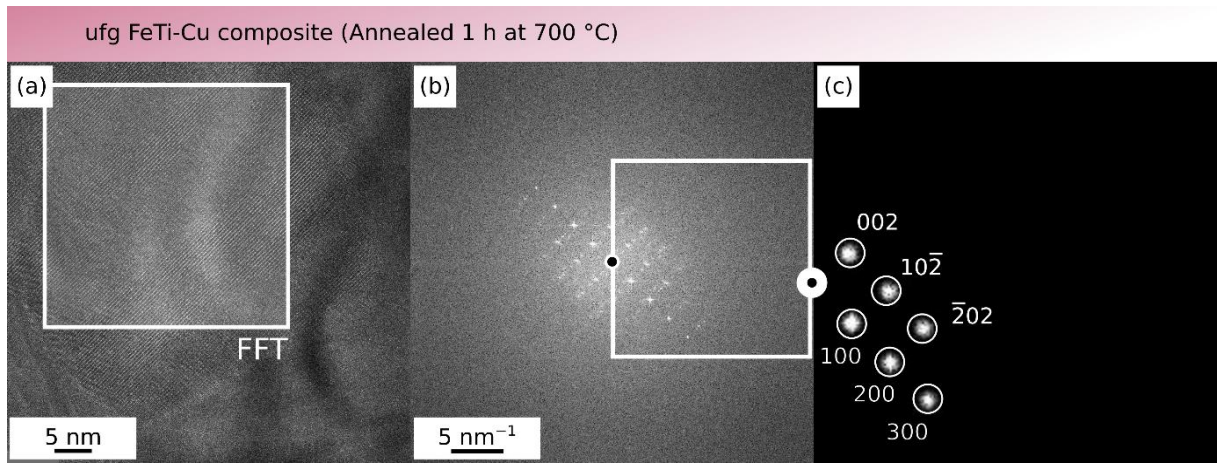

**Figure S9:** (a) HRTEM micrograph and (b) the derived and (c) indexed fast Fourier transformation (FFT) of a Fe$_2$T precipitate. Measurements were made on a ufg FeTi−Cu composite.

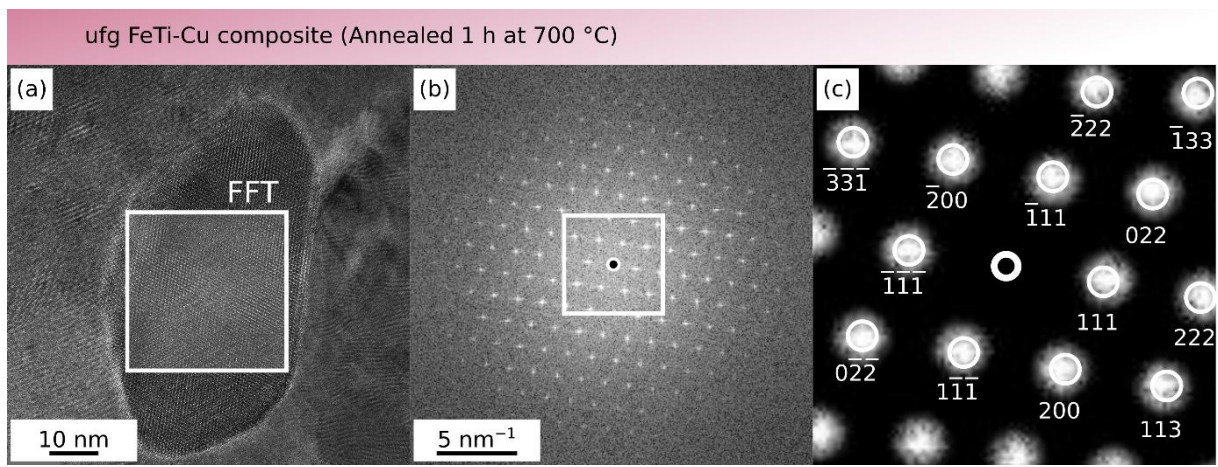

**Figure S10:** (a) HRTEM micrograph and (b) the derived and (c) indexed FFT of a Ti$_2$Fe or Ti$_4$Fe$_2$O precipitate. Measurements were made on a ufg FeTi−Cu composite.

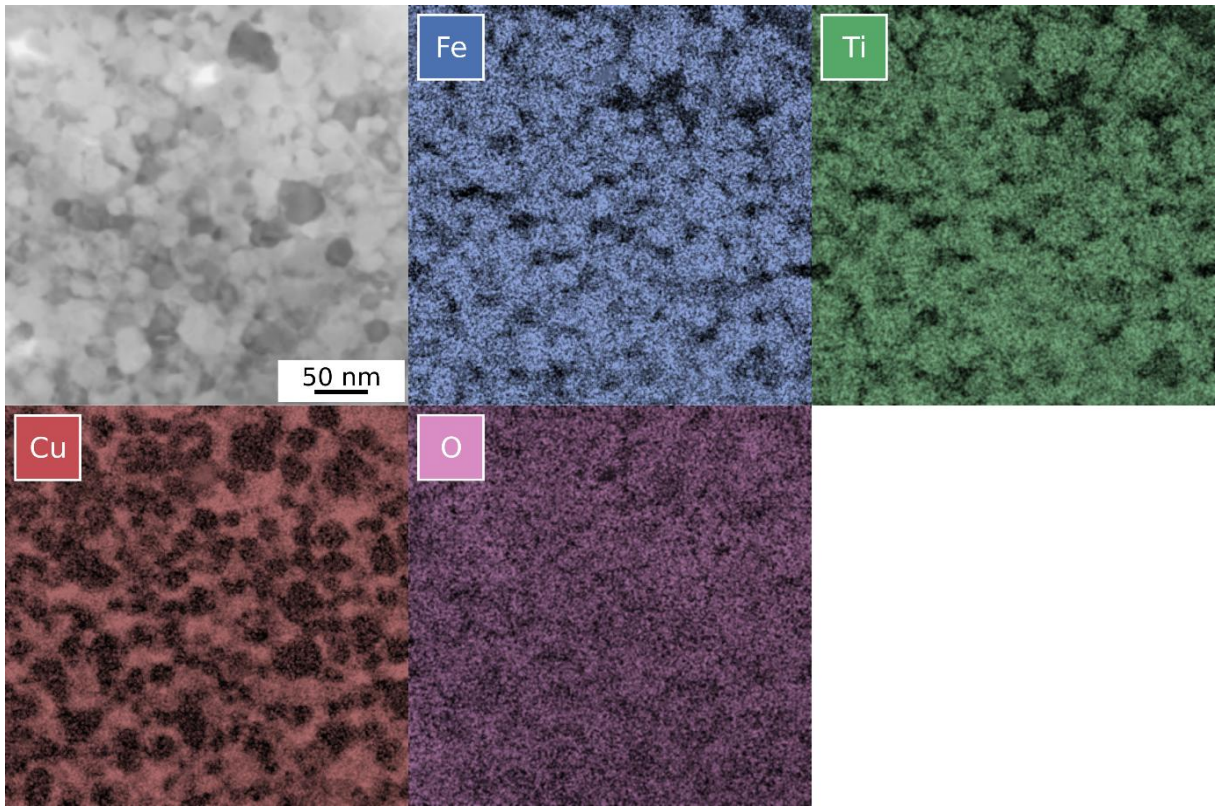

**Figure S11:** TEM-EDX maps of nc FeTi−Cu composites.

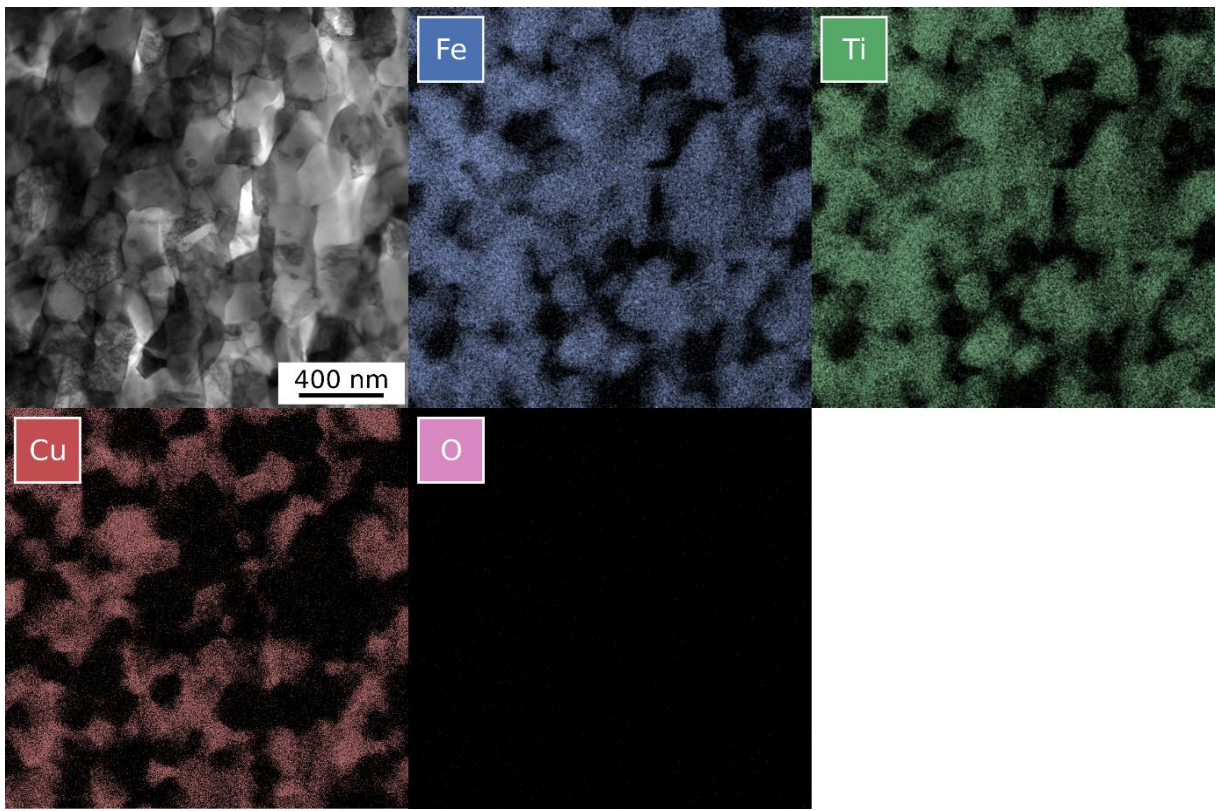

**Figure S12:** TEM-EDX maps of ufg FeTi−Cu composites.

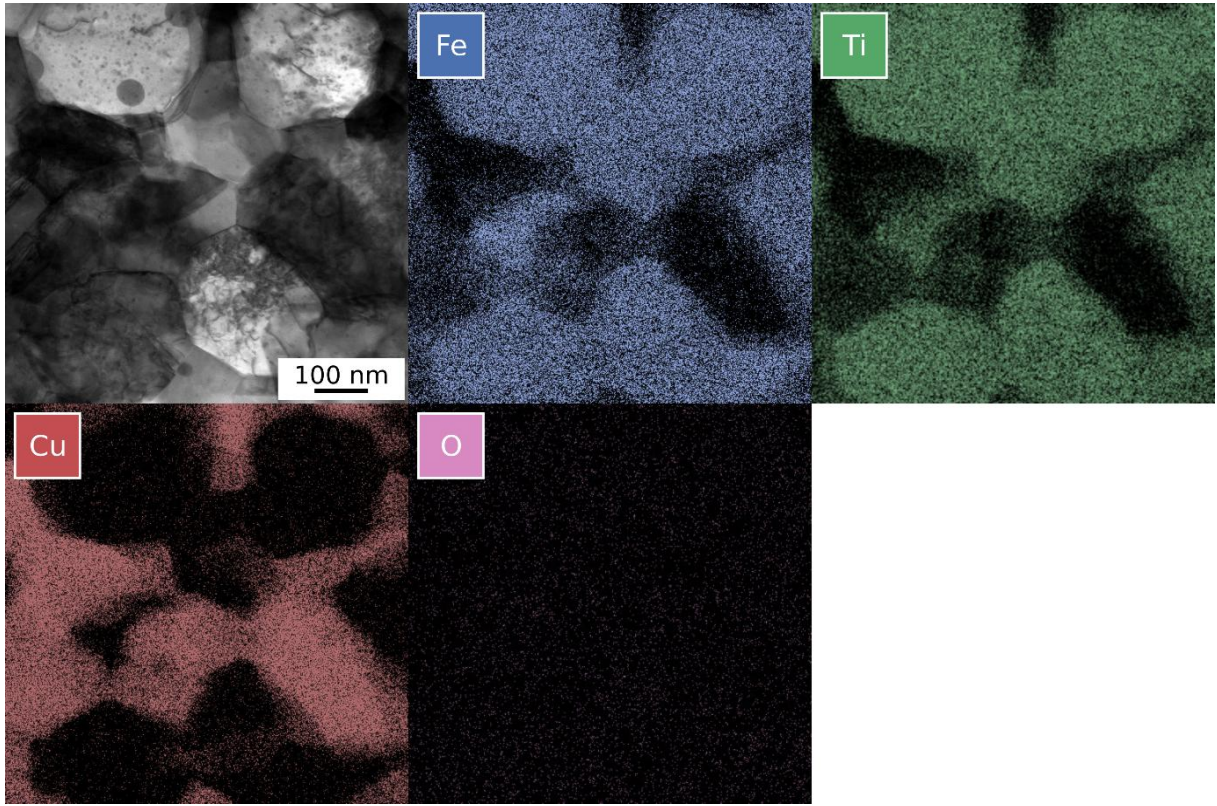

**Figure S13:** TEM-EDX maps of ufg FeTi−Cu composites. This Figure was taken at a higher magnification compared to **Figure S12.**

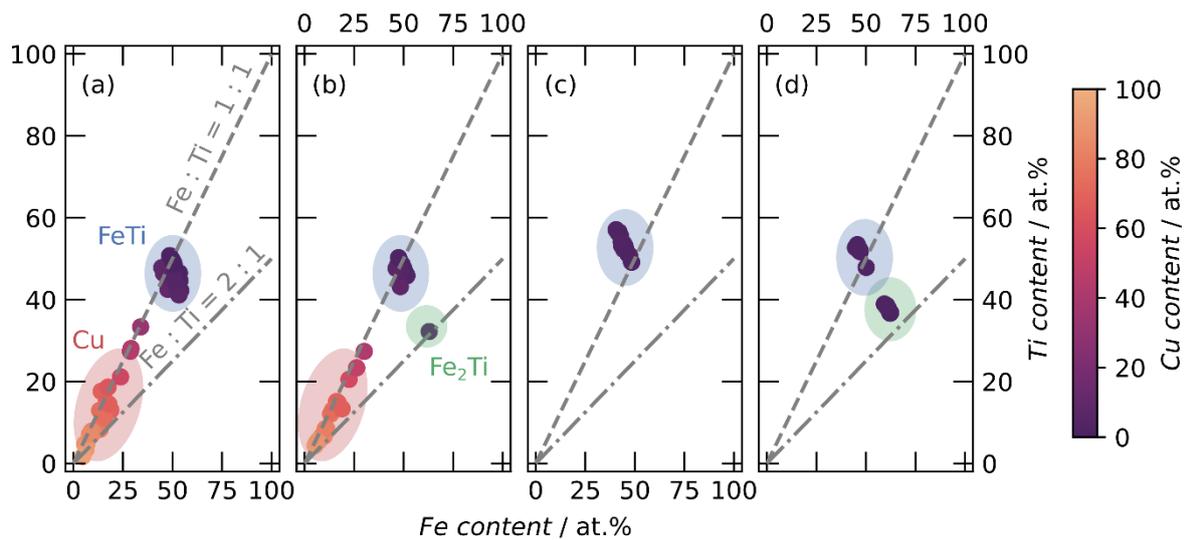

**Figure S14** Elemental composition determined via TEM EDX of the (a) nc and (b) ufg FeTi−Cu composites, as well as the derived (c) np and (d) ufp FeTi.

# FeTi−Cu composites: Etching process

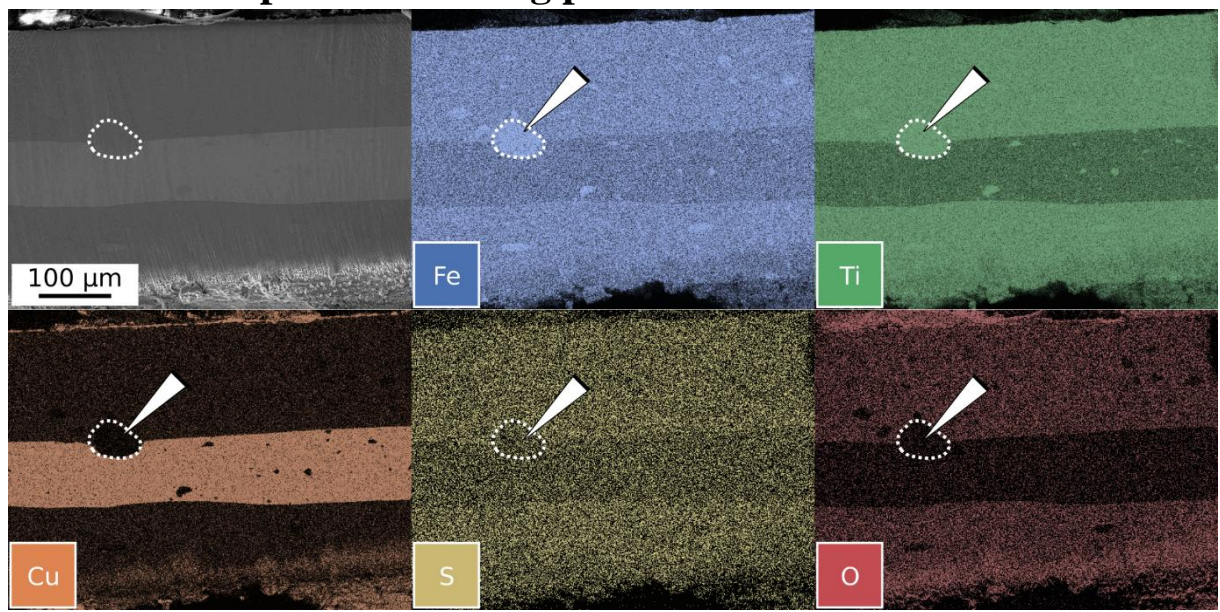

**Figure S15**: EDX maps of nc FeTi−Cu composites etched for 3 h in 1 M ammonium persulfate solution. White arrows indicate larger, residual FeTi particles.

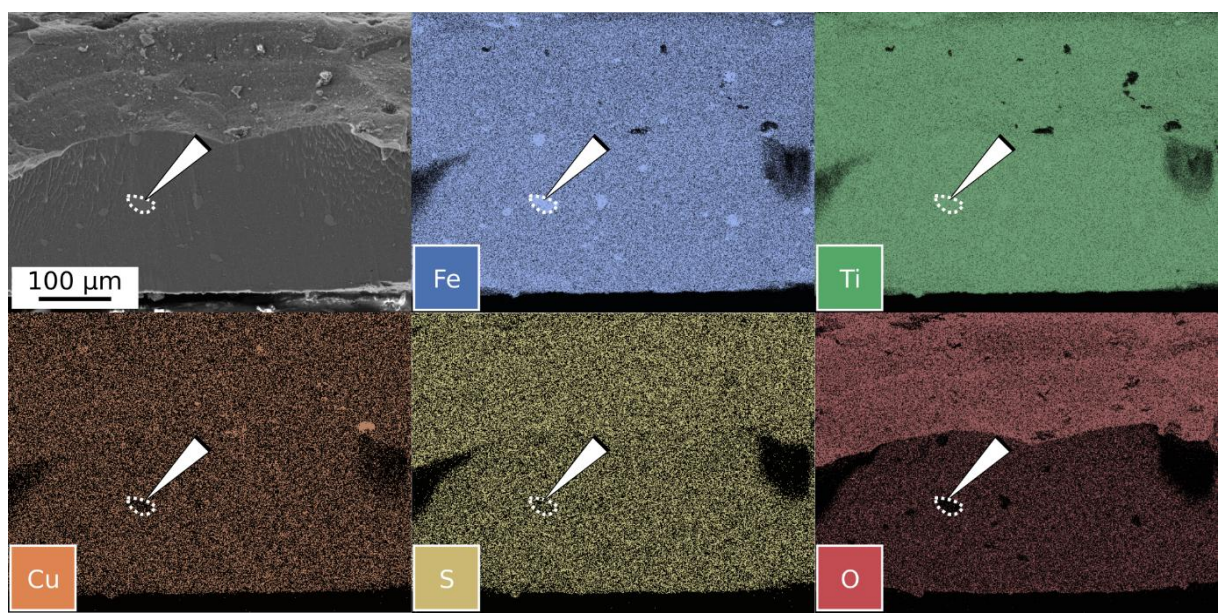

**Figure S16**: EDX maps of nc FeTi−Cu composites etched for 24 h in 1 M ammonium persulfate solution. White arrows indicate larger, residual FeTi particles.

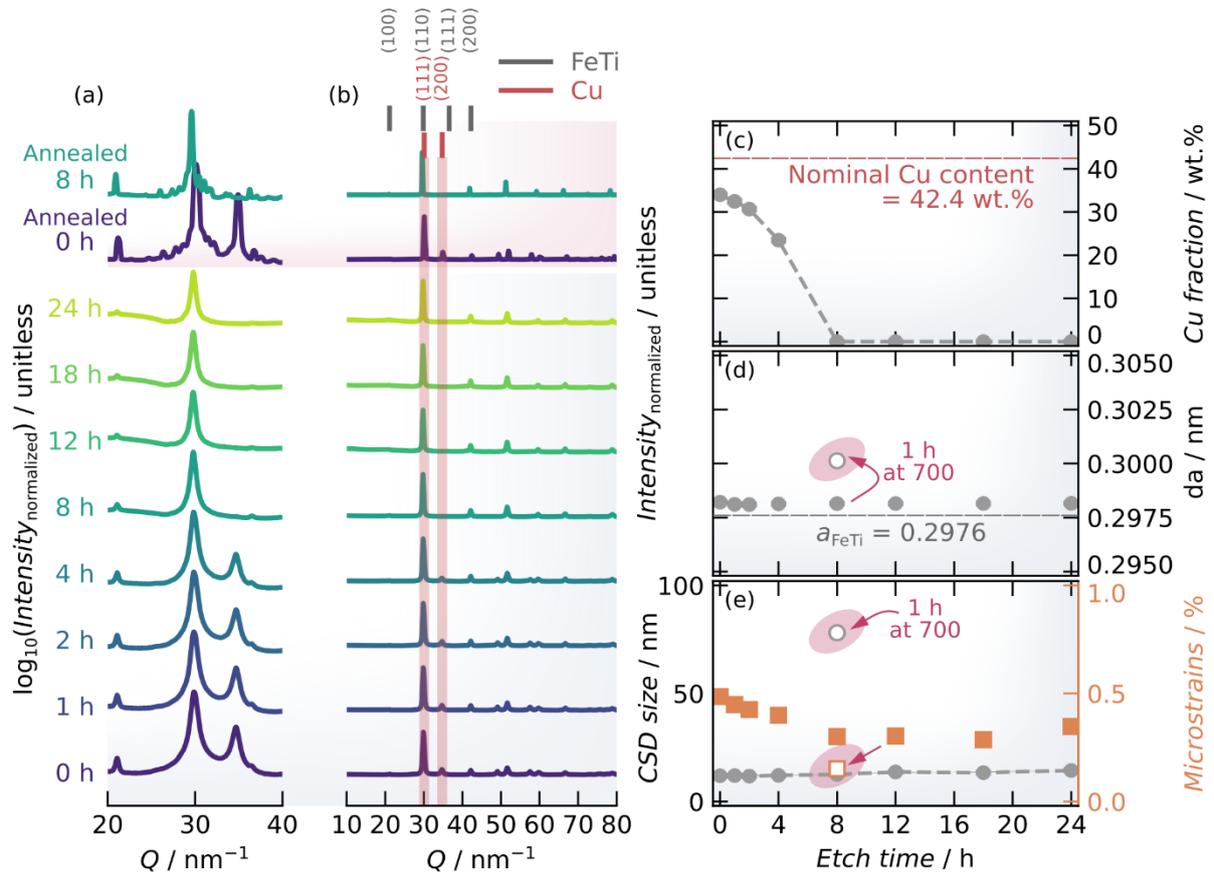

**Figure S17:** SR-XRD patterns (normalized to the maximum peak intensity of the respective pattern) of nc and ufg FeTi−Cu composites etched for different durations. SR-XRD patterns are plotted (a) logarithmically and (b) linearly. Variations in (c) Cu phase fraction, (d) lattice constant, (e) microstrain / CSD size derived from Rietveld refinements. Open symbols represent values obtained from ufg FeTi−Cu composites and are highlighted by arrows.

Rietveld refinement results shown in **Figure S17 (c-e)** confirmed the reduction and subsequent disappearance of the Cu phase with etching. The lattice constant is slightly increased compared to the reference value of FeTi, but remains relatively stable with etch time. A slight increase aligns with Cu substitution within FeTi due to HPT-induced mechanical alloying, leading to a lattice expansion. The coherent scattering domain (CSD) size remains relatively stable, but microstrains are slightly lowered during etching. The latter relaxation could be introduced by removing a constituent phase of the nanocomposite, thereby allowing the residual material a greater degree of freedom to accommodate residual stresses. In line with all the previous investigations, the annealed sample has a significantly higher CSD size, albeit lower compared to the results from electron microscopy.

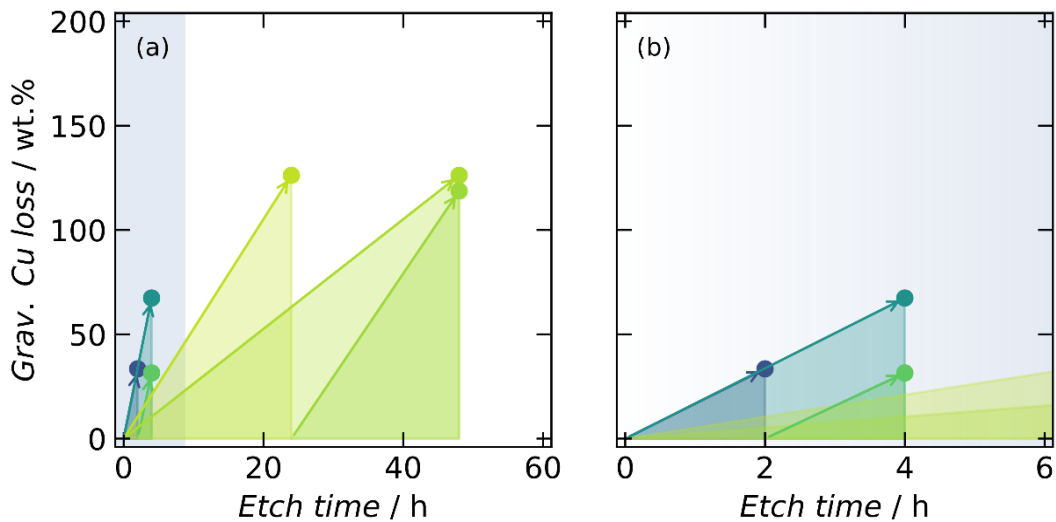

**Figure S18**: (a) Results of the planned interval tests on nc FeTi−Cu composites immersed in 1 M $(NH_4)_2SO_8$ solution at RT. (b) the zoomed-in region at shorter etch times.

To investigate whether the changing etch rates are related to changes in the material or the etchant over time, so-called planned interval tests were performed. The results are shown in Figure S18. Two nc FeTi−Cu composites were immersed in freshly prepared 1 M ammonium persulfate solution. After an "interval" of 2 h, one sample was removed from the solution, and a new sample was immersed in the now already used etchant. The etching was continued for another 2 h, after which both samples were removed from the solution. The same procedure was also done with an interval of 24 h. The results indicate that the reduced etch rate originates from the sample itself, rather than from the potential decomposition/consumption of the ammonium persulfate etchant, as the mass losses of the individual intervals are similar for both 2-hour and 24-hour intervals.

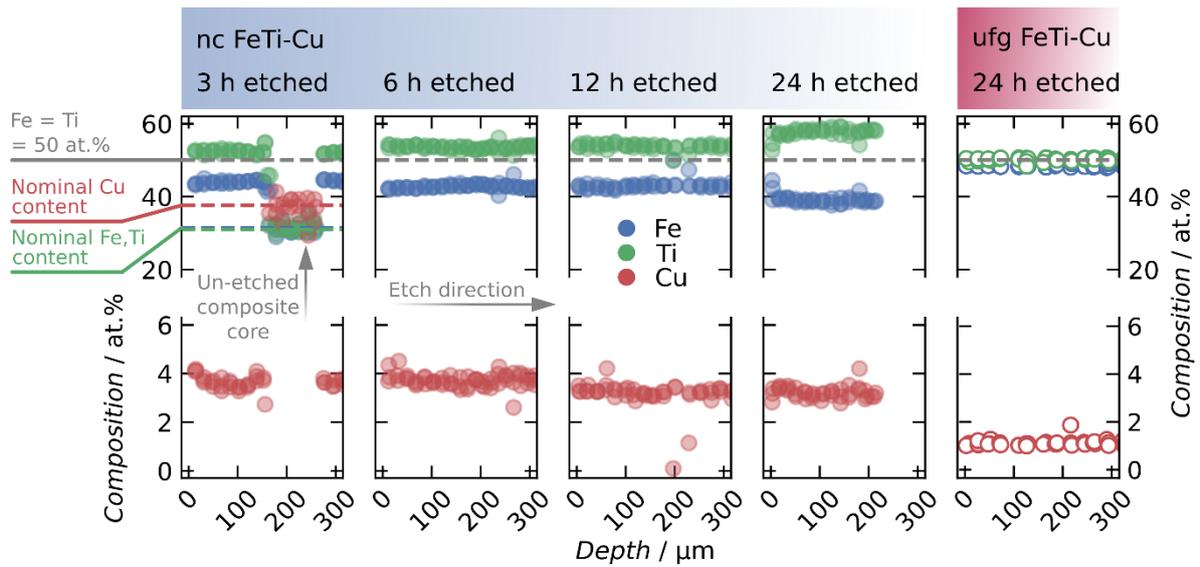

**Figure S19:** SEM EDX results for both nc (np) and ufg (ufp) FeTi−Cu composites subjected to etching. The measurements were performed along the axial (etching) direction on the cross-section of the HPT disks. The results are provided as a function of (etching) depth and time.

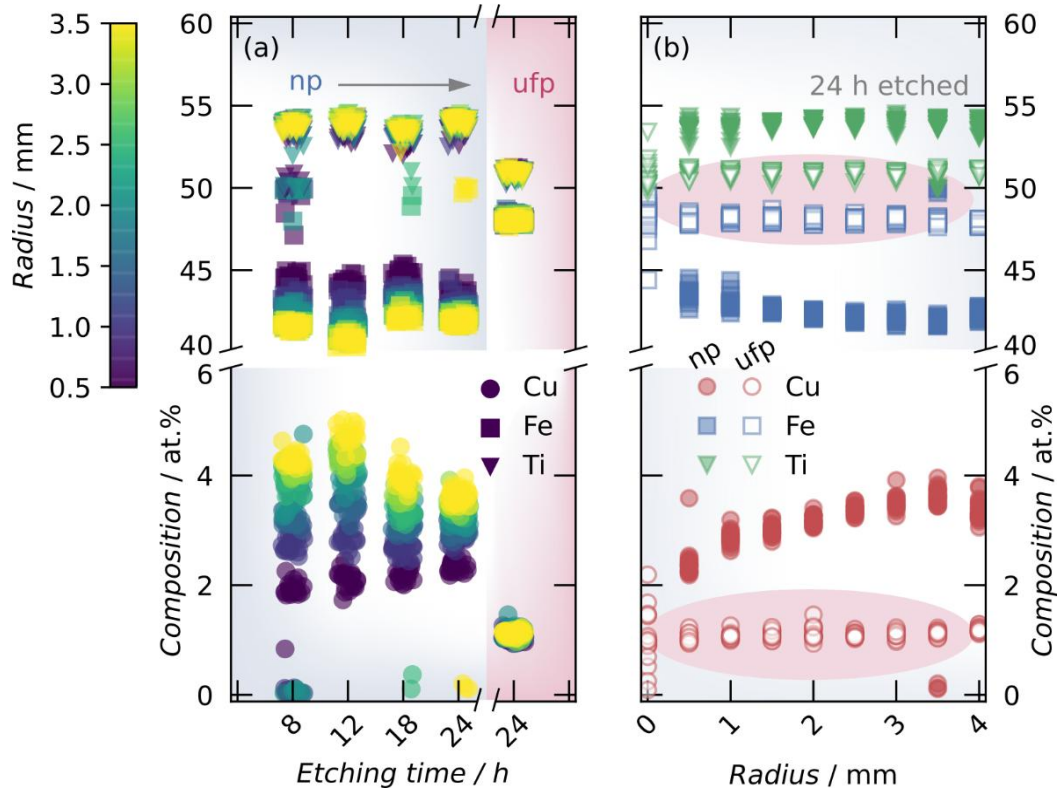

**Figure S20:** SEM EDX results for both the nc (np) and ufg (ufp) FeTi−Cu composites subjected to etching. (a) The elemental composition as a function of etching time, with coloration varying according to radial position. (b) Elemental composition as a function of radial composition for samples subjected to 24 h etching. Open symbols represent values obtained from ufg FeTi−Cu composites and are highlighted by arrows.

**Table S3**: Compositions of the FeTi phase before and after etching obtained by TEM EDX, and SEM EDX.

|  |  | *Fe* / **at.%** | *Ti* / **at.%** | *Cu* / **at.%** |
|---|---|---|---|---|
| **as-HPT** | Unetched composite (TEM) | 50.7±2.5 | 45.0±1.8 | 4.3±2.2 |
|  | Etched – 24 h (SEM) | 42.3±1.3 | 53.7±0.7 | 3.4±0.6 |
|  | Etched – 40 min (TEM) | 44.6±1.7 | 52.0±1.7 | 3.4±1.1 |
| **Annealed** | Unetched composite (TEM) | 50.2±0.8 | 46.9± 0.9 | 3.0±0.7 |
|  | Etched – 24 h (SEM) | 48.1±0.2 | 50.8±0.2 | 1.1±0.1 |
|  | Etched – 40 min (TEM) | 46.4±1.1 | 52.1±1.1 | 1.3±0.4 |

## Porous FeTi: Structural characterization

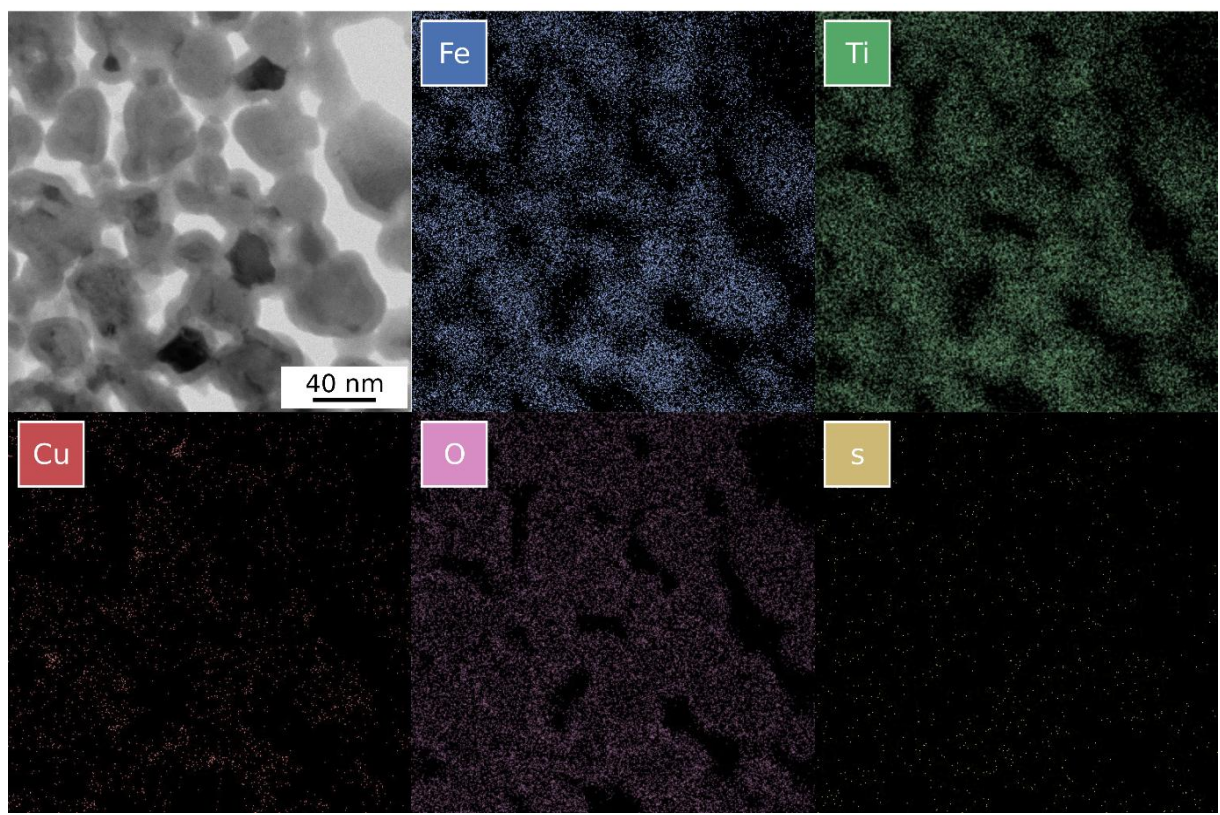

**Figure S21:** TEM EDX maps of np FeTi. The TEM sample of the nc FeTi−Cu composite was immersed for 40 min in an 1 M ammoniumpersulfate solution.

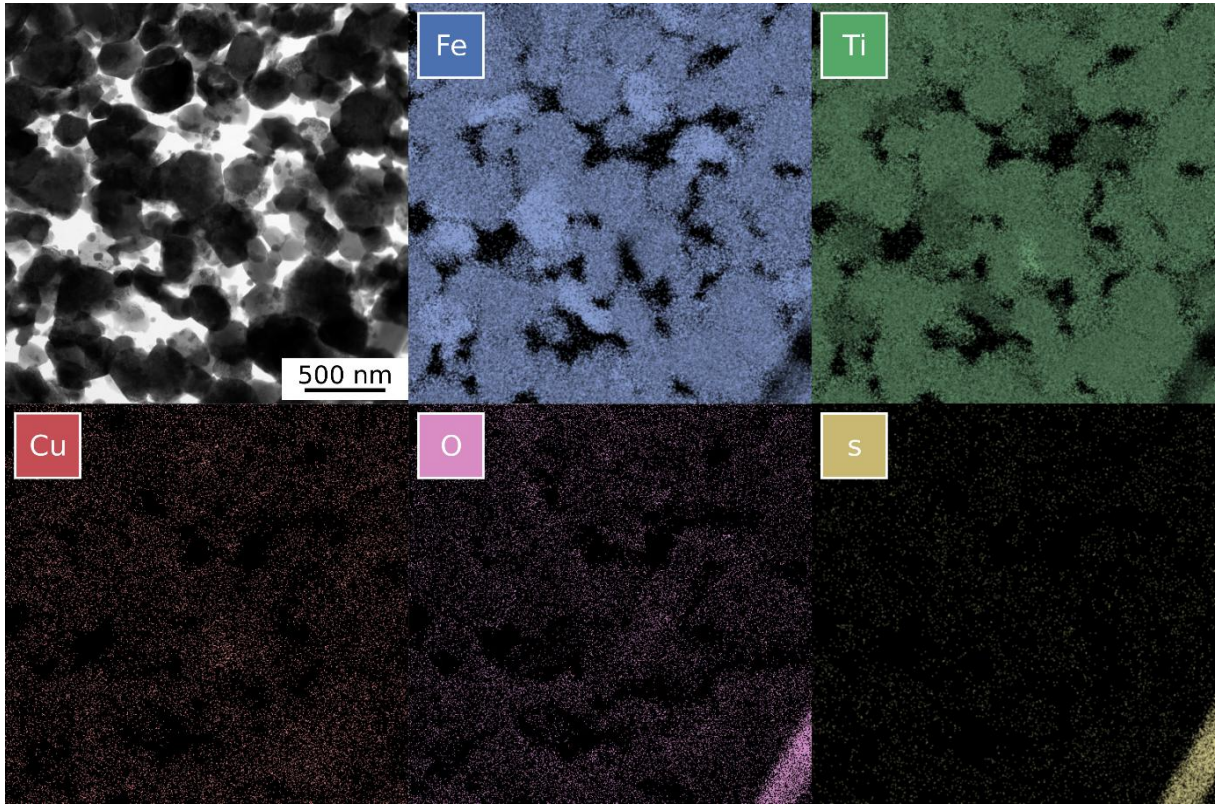

**Figure S22**: TEM EDX maps of ufp FeTi. The TEM sample of the ufg FeTi−Cu composite was immersed for 40 min in an 1 M ammoniumpersulfate solution.

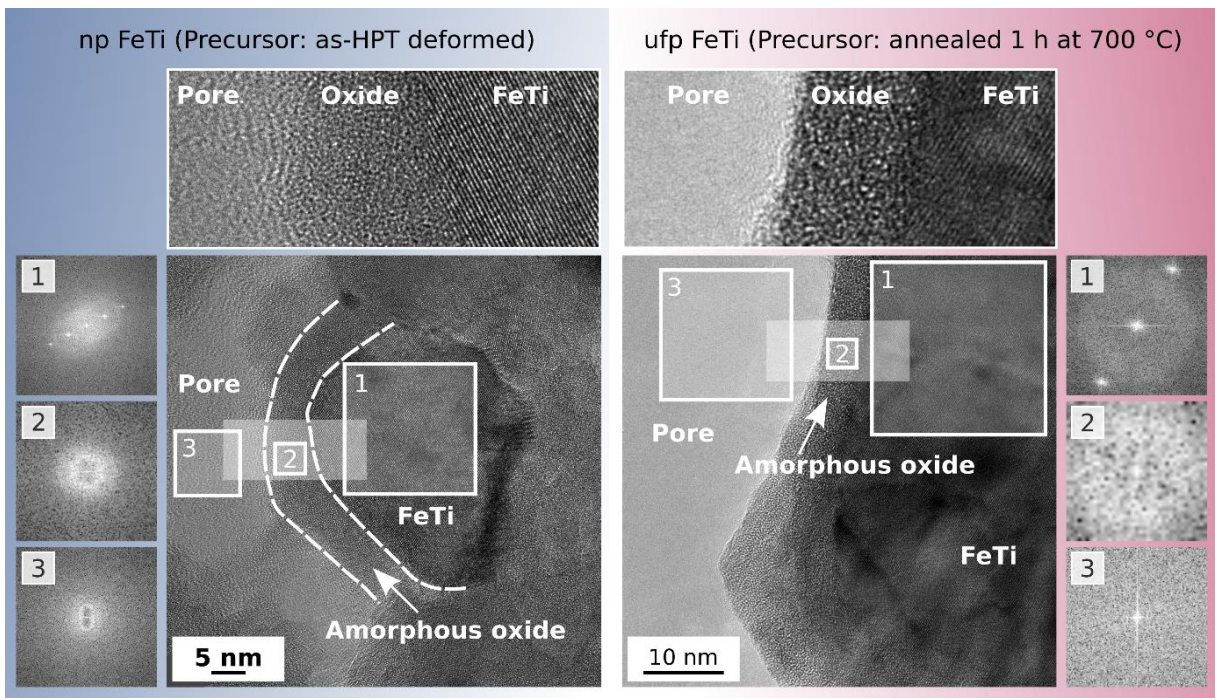

**Figure S23**: HR-TEM micrographs of (a) np and (b) ufp FeTi showing the thickness and amorphous nature of the oxide layers covering the FeTi ligaments.

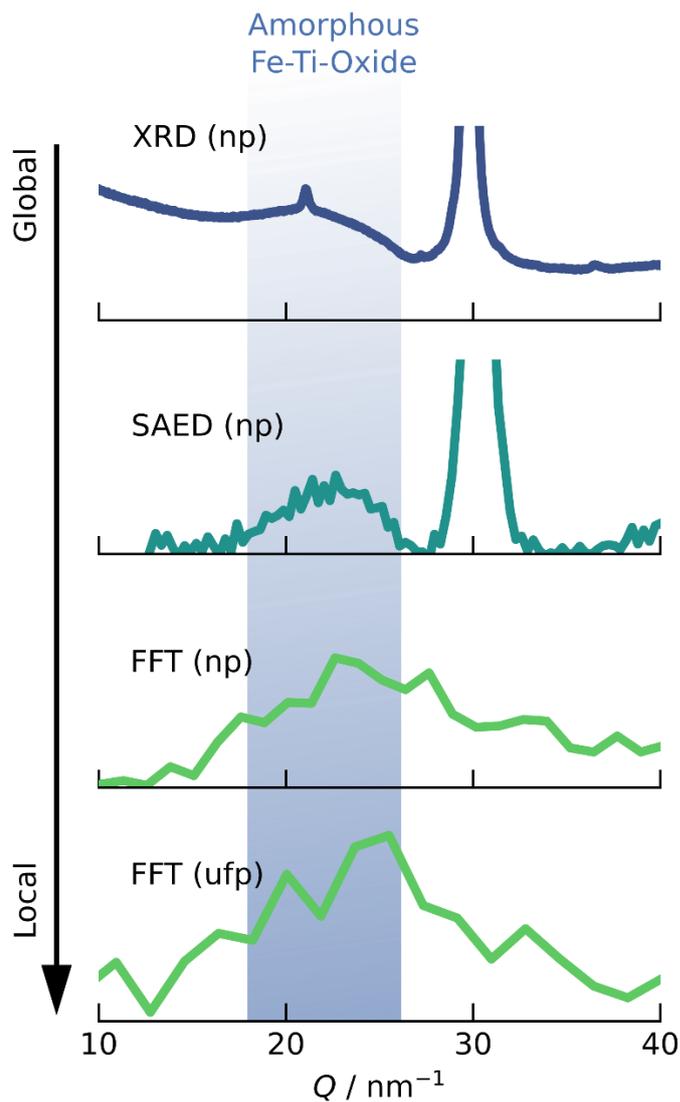

**Figure S24**: (a) XRD, (b) SAED, and (c) FFT of HRTEM micrographs highlighting the amorphous structure of the oxide covering the ligaments in np and ufp FeTi after etching.

**Figure S24** compares XRD and TEM results. A broad hump is clearly visible at about 22 nm$^{-1}$ in both the XRD and SAED patterns. FFT of the oxide layers in the HRTEM micrographs, as seen in **Figure 23**, and subsequent integration also yields peaks in the same region. It should be noted that the region subjected to FFT was rather small, and the resulting uncertainty could explain certain deviations from the XRD and SAED data.

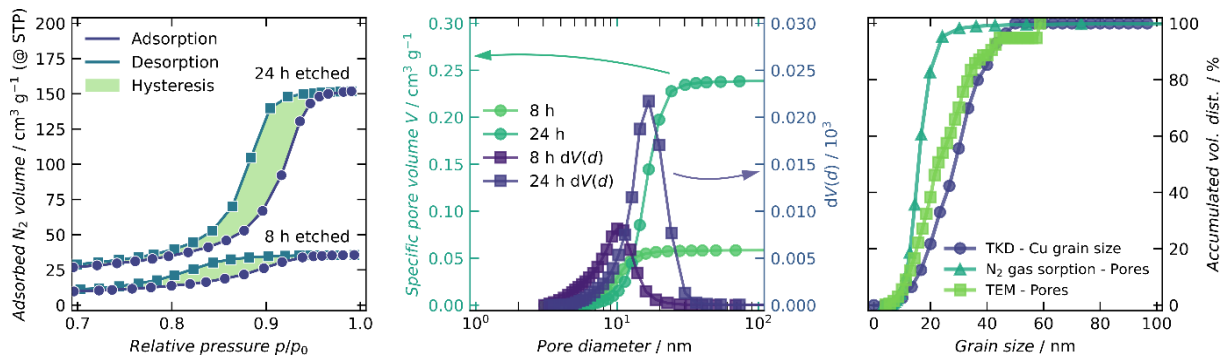

**Figure S25:** (a) The isotherms recorded during $N_2$ gas adsorption and desorption at 77 K on np FeTi and (b) the corresponding pore size distributions and cumulative pore volumes. (c) Comparison of the cumulative pore / Cu volumes obtained using various direct and indirect characterization methods.

$N_2$ gas adsorption and desorption measurements allowed the determination of the specific surface area using the Brunauer-Emmett-Teller (BET) method and the pore size distribution using the Barrett-Joyner-Halenda (BJH) method. The Cu grains of the nanocomposite, mapped by TKD, were compared to the pores, as Cu grains can be regarded as pore precursors. Consequently, these results show that Cu dissolution of the nc FeTi−Cu composite yields an isotropic, mesoporous material with a relatively high specific surface area and well-defined pore size distribution. The annealed material has too large pores (beyond nanopores) for proper $N_2$ gas sorption analysis. The pore fraction was determined by comparing the measured pore volume, the sample mass, and the sample bulk density. The latter was determined using He Pycnometry (Helium Pyknometer Ultrapyc 5000, Anton Paar).

**Table S4**: Results obtained from the $N_2$ gas adsorption/desorption data collected at 77 K.

| *Etch time /* h | *Specific surface area (BET)* / $m^2\ g^{-1}$ | *Pore volume /* $cm^3\ g^{-1}$ | *Pore fraction /* unitless | *Pore size (BJH) /* nm |
| --- | --- | --- | --- | --- |
| 8 | 16.5 | 0.05 | 25.0 | 10.0 |
| 24 | 48.6 | 0.23 | 57.2 | 16.7 |

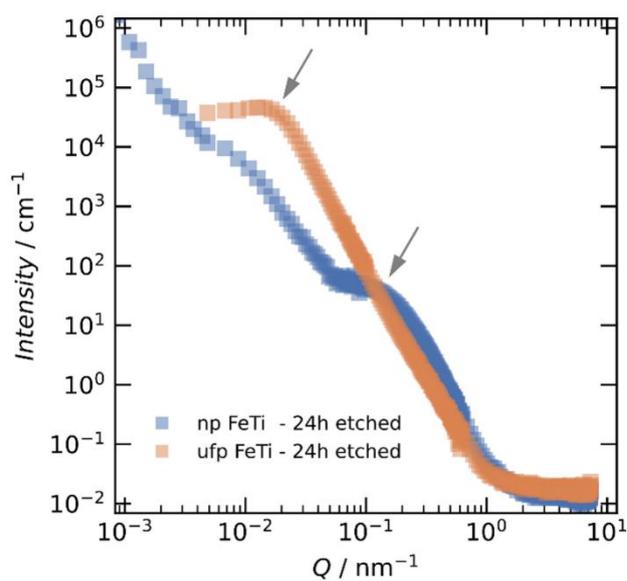

**Figure S26:** Plot showing the SANS and USANS data obtained from np and ufp FeTi, with arrows indicating the ligament sizes as determined by the relation $D \approx Q/2\pi$.

## Porous FeTi: Hydrogen absorption/desorption properties

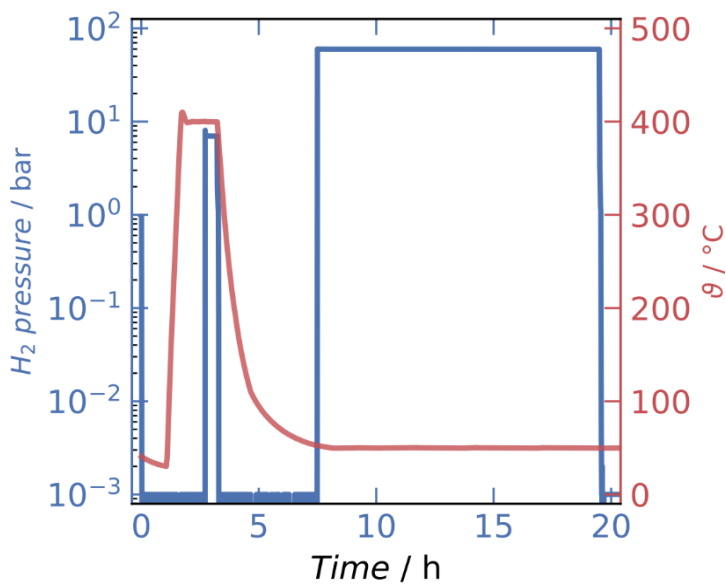

**Figure S27:** Plot showing the pressure and temperature ($\vartheta$) during the FeTi activation procedure. The program follows the procedure proposed by Reilly and Wiswall.[3]

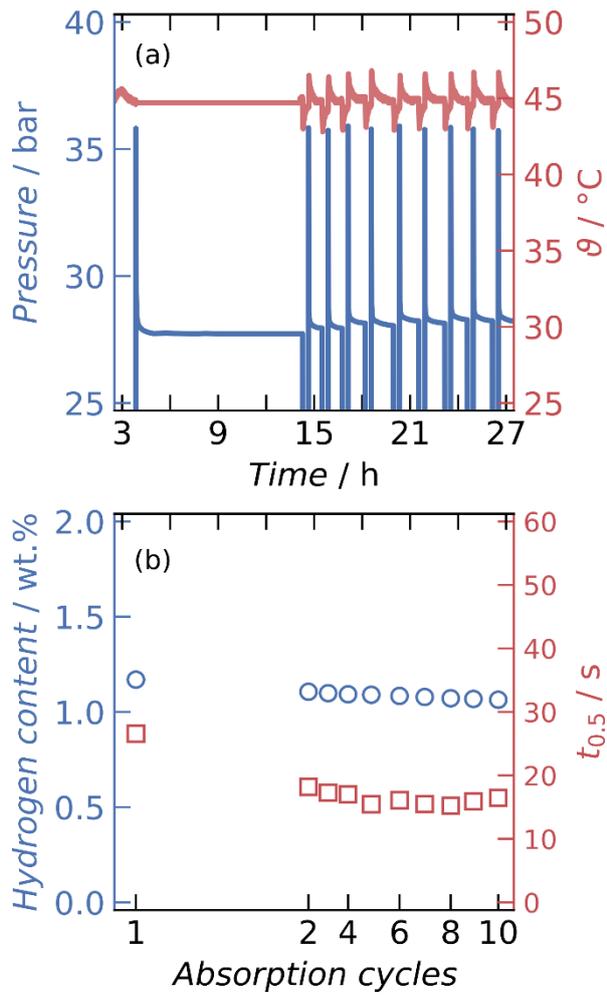

**Figure S28:** Absorption and desorption cycling of ufp FeTi. In (a), the pressure and temperature ($\vartheta$) conditions during cycling and in (b) the obtained capacities and times until 50 % capacity ($t_{0.5}$) are shown.

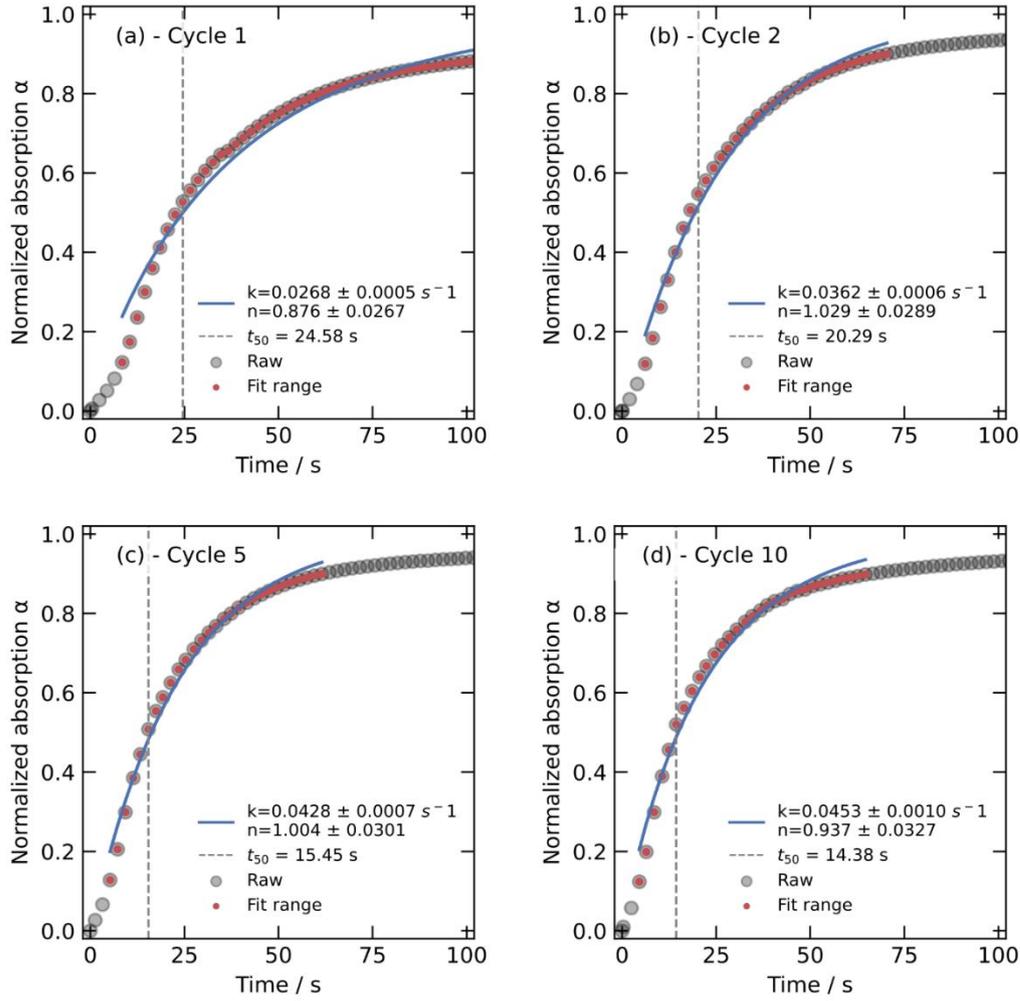

**Figure S29:** Results of fitting the Johnson-Mehl-Avrami-Kolograow kinetic model to the absorption data. Exemplary curves of the (a) 1st, (b) 2nd, (c) 5th, and (d) 10th cycle are given.

Different kinetic models were fitted (JMAK: n = 1, 0.5, 0.33, 0.25, 0.4; geometrical contraction models - interface controlled: 2D & 3D; diffusion models: 1D, 2D, 3D).[4]¶ The best fit was achieved for the Johnson-Mehl-Avrami-Kolograow model with $n = 1$:

$$\alpha = 1 - \exp(-(k\,t)^n), \tag{S1}$$

With $\alpha$ being the reacted fraction, $k$ the rate constant, and $n$ the dimensional factor (i.e., $n = 1$ for 1D growth). As seen in **Figure S29**, subsequent fitting of the JMAK with both $k$ and $n$ as fitting parameters still resulted in a good fit with $n \approx 1$.

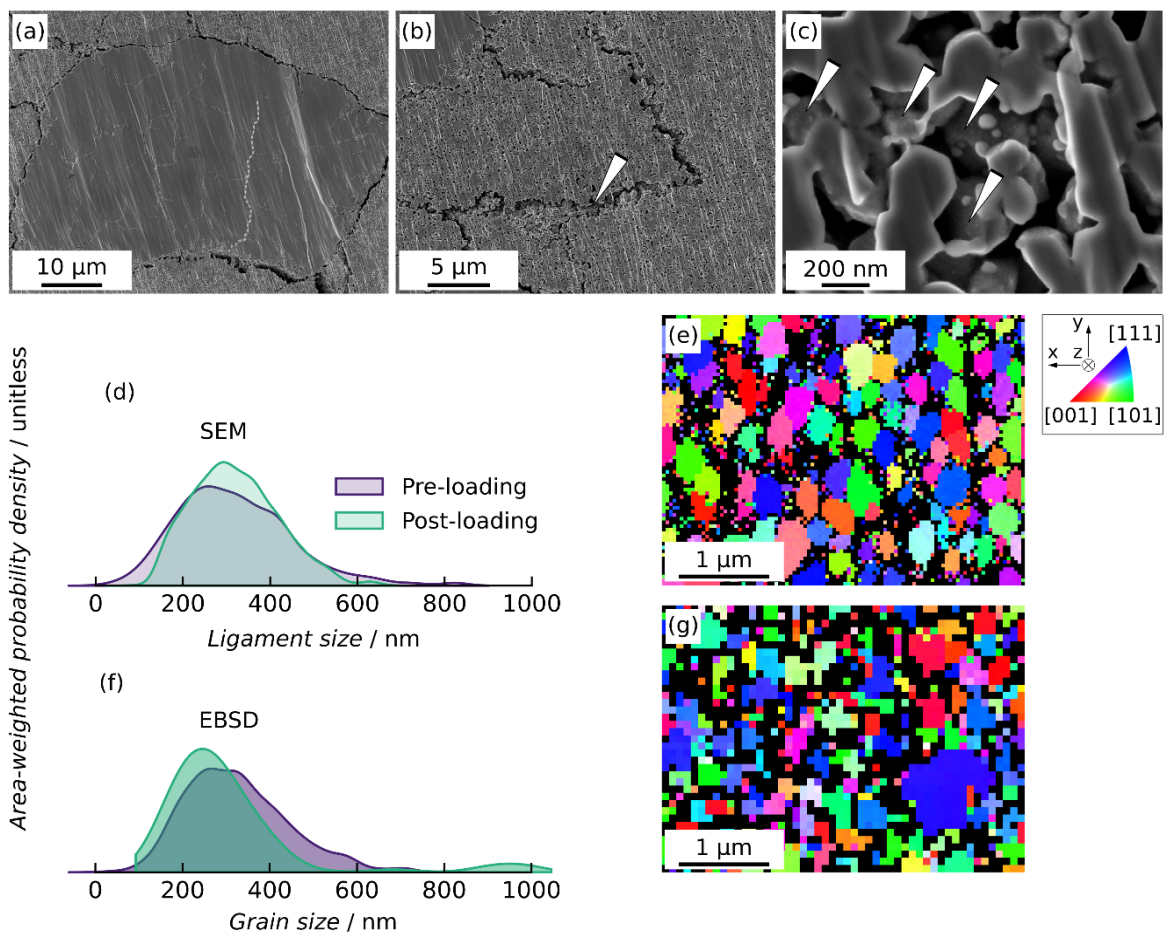

**Figure S30:** Post sorption characterization (after the measurement of the PCI curves and 10 absorption/desorption cycles) of ufp FeTi. (a-c) SEM micrographs of a particle cross-section. Ligament and grain size distributions of (d) pristine and (f) cycled ufp FeTi obtained by SEM and EBSD, respectively. IPF maps obtained by EBSD of (e) pristine and (g) cycled ufp FeTi.

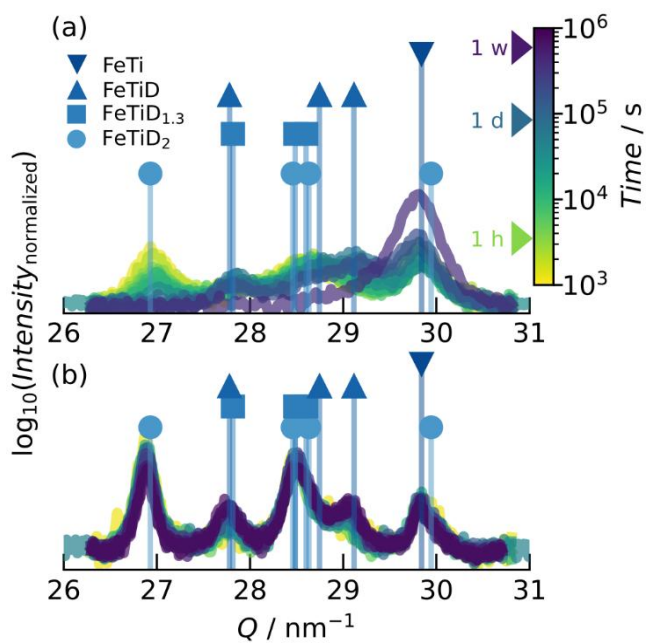

**Figure S31:** XRD recorded at specific times following (potentiostatic) electrochemical loading of (a) np and (b) ufp FeTi in 0.1 M $H_2SO_4$ at −600 mV. An Ag/AgCl reference electrode was used.

The enhanced permeability of the oxide covering the np FeTi was confirmed by (potentiostatic) electrochemical loading at −0.6 V vs. Ag/AgCl in 1% $H_2SO_4$. As visible in **Figure S31**, the XRD patterns recorded immediately following hydrogen loading show the formation of hydrides for both np and ufp FeTi. The discrepancy to the gas loading could be due to the aggressive nature of the $H_2SO_4$ electrolyte, potentially attacking the covering oxide layer, as well as the enormous (electric potential-induced) driving force for hydrogenation. Regarding the latter, converting the electrochemical potential to an equilibrium gas pressure yields a value of >$10^6$ bar. Nevertheless, np FeTi showed rather fast desorption and decomposition of the hydride, while the ufp counterpart remained stable for extended periods, as expected for "oxide-stabilized" FeTi.[3] This indicated a drastically higher permeability of np FeTi.

# Analytical core-shell (Tessier) model

In a metal-hydrogen system, the hydrogen dissolved at the initial sites and the hydrogen gas are in thermodynamic equilibrium and are therefore at the same chemical potential. The equation describing this is

$$\frac{1}{2} k_b T \ln\left(\frac{p}{p_0}\right) = \mu_0(T) - \frac{E_d}{2} + k_b T \ln\left(\frac{c}{\frac{N_H}{N_L} - c}\right) - a_{int} c + \Delta\mu, \tag{S2}$$

with $p$ being the hydrogen pressure, $k_b$ the Boltzmann constant, and $T$ the temperature in K. The first term on the right is $\mu_0$; the chemical potential at infinitesimal concentrations; the second term introduces the dissociation energy $E_d$ required to split the $H_2$ molecule into atomic H; the third term describes the configurational entropy of the H atoms distributed over the metal intersitial sites, with $c$ being the hydrogen concentration in the metal, $N_H$ the number of hydrogen sites, $N_L$ the number of metal/lattice sites. The fourth term, $a_{int}\, c$, describes the concentration dependency of the chemical potential. The fifth and last term, $\Delta\mu$, includes the microstructure-dependent elastic stresses induced by hydrogen absorption. This term was separated from the general term $a_{int}\, c$ due to its significance in this work. Specifically, this term constitutes the impact of the FeTi−oxide core-shell structure on hydrogen absorption and subsequent hydride formation. It will be derived in the following using the procedure proposed by Tessier et al..[5]

Wagner and Horner defined the force dipole tensor as the stresses induced per hydrogen atom. In a simple and isotropic model by Tessier, this force dipole tensor $P$ is approximated by

$$P = -B\, v_H, \tag{S3}$$

with B being the bulk modulus (for FeTi assumed as $\approx 192$ GPa)[6] and $v_H$ the volume expansion per hydrogen atom (for FeTi assumed as $\approx 2.8$ Å$^3$)[5,7,8]. Summing up over all hydrogen atoms yields the density of the force dipole tensor $\Pi$, being the stress tensor describing the stresses at position $x$ induced by the partial density of hydrogen $\rho_j$ on the sites of type $j$

$$\Pi_{\alpha\beta}(x) = \sum_j \rho_j(x) P_{\alpha\beta j}. \tag{S4}$$

Considering **Equation S3** and assuming a homogenous and isotropic hydrogen distribution, yields a simplified equation for the hydrogen introduced stress,

$$\Pi = c_H\, P = \frac{N_H}{V} P, \tag{S5}$$

with $c_H$ being the hydrogen concentrations, $N_H$ the number of hydrogen sites, and $V$ the overall material volume; $P$ is the isotropic force dipole tensor. These stresses result in a response of the host lattice (harmonic response), and the following equation needs to be balanced to be in mechanical equilibrium,

$$\sigma_{\alpha\beta}(x) = \Pi_{\alpha\beta}(x) + C_{\alpha\beta\mu\nu}\varepsilon_{\mu\nu}(x), \tag{S6}$$

with $\sigma_{\alpha\beta}(x)$ being the external or net stress, $\Pi$ the density of the fore-dipole tensor as defined in **Equations S4 and S5**, respectively; $C_{\alpha\beta\mu\nu}$ is the elastic tensor of the hydride-forming material, and $\varepsilon_{\mu\nu}$ the associated strain. Wagner and Horner[9] set the net stress to zero. However, Tessier et al.[5] and this study keep it because of the confinement in the core–shell structure.

Inserting **Equation S5** in **S6**, assuming an isotropic and hydrostatic backpressure $\sigma$ (instead of $\sigma_{\alpha\beta}(x)$), as well as an isotropic compressibility $K_c$ (instead of the elastic tensor $C_{\alpha\beta\mu\nu}$ or the compliance tensor $S_{\mu\nu\alpha\beta}$, respectively), the strains in the crystalline core $\varepsilon_{c,\mu\mu}$, and the oxide shell $\varepsilon_{s,\theta\theta}$ (the subscripts indicate *core* and *shell*, respectively) are given by

$$\varepsilon_{c,\mu\mu} = -S_{c,\mu\mu\alpha\beta}\left(\Pi_{c,\alpha\beta} + \sigma_{c,\alpha\beta}\right) = -\frac{K_c N_c P_c}{V_c} + K_c\,\sigma_c \;, \tag{S7}$$

and,

$$\varepsilon_{s,\theta\theta} = -\frac{1}{3}\frac{K_s N_s P_s}{V_s} + S_{s,\theta\theta\alpha\alpha}\sigma_{c,\alpha\alpha} \;. \tag{S8}$$

To relate these strains, the necessary boundary condition is that the core and shell will remain in contact, which is ensured by the condition,

$$\varepsilon_{s,\theta\theta} = \frac{1}{3}\varepsilon_{c,\mu\mu} \;, \tag{S9}$$

with $\varepsilon_{s,\theta\theta}$ being the strain on the (oxide) shell, while $\varepsilon_{c,\mu\mu}$ is the strain in the (FeTi) core. With $N_c$ and $N_s$ being the numbers of hydrogen atoms in the core and shell, $V_c$ and $V_s$ the core and shell volumes, and $P_c$ and $P_s$ the respective dipole-force tensors. $S_{s,\theta\theta\alpha\alpha}$, and $\sigma_{s,\alpha\alpha}$ are the compliance and stress tensors describing the oxide shell. We assume that the original, amorphous oxide layer does not absorb appreciable amounts of hydrogen, therefore setting the first term ($-\frac{1}{3}\frac{K_s N_s P_s}{V_s}$) in **Equation S8** to zero. Inserting **Equations S7 and S8** into **S9** yields

$$-\frac{K_c N_c P_c}{V_c} + K_c\,\sigma_c = -\frac{K_s N_s P_s}{V_s} + 3\,S_{s,\theta\theta\alpha\alpha}\sigma_{s,\alpha\alpha} \;. \tag{S10}$$

$S_{s,\theta\theta\alpha\alpha}$ can again be assumed to be isotropic; however, we do not yet have a simplified term for the shell stress tensor $\sigma_{\alpha\alpha}$. The stress components $\sigma_{rr}$, $\sigma_{\theta\theta}$, and $\sigma_{\phi\phi}$ (in spherical coordinates) in a core-shell structure at a radius $r$ are related to the internal and hydrostatic pressure in the shell by the equations[5]

$$\sigma_{rr} = -\frac{\sigma R_1^3}{R_2^3 - R_1^3}\left(1 - \frac{R_2^3}{r^3}\right), \qquad (S11)$$

$$\sigma_{\theta\theta} = \sigma_{\phi\phi} = -\frac{\sigma R_1^3}{R_2^3 - R_1^3}\left(1 + \frac{R_2^3}{2r^3}\right), \qquad (S12)$$

with $R_1$ and $R_2$ being the internal and external radii of the shell, respectively. Since the stresses and strains at the contact between core and shell are the determining factor, we assumed $r=R_1$. The compliance tensor for the isotropic, homogeneous material was assumed to be [10]

$$S_s = \frac{1}{Y_a}\begin{bmatrix} 1 & -\nu_a & -\nu_a \\ -\nu_a & 1 & -\nu_a \\ -\nu_a & -\nu_a & 1 \end{bmatrix}. \qquad (S13)$$

Solving the last term $3\,S_{s,\theta\theta\alpha\alpha}\sigma_{s,\alpha\alpha}$ in **Equation S10** using **Equations S11-S13** and setting r =$R_1$ yields an expression for the backpressure in the core as a function of the hydrogen content,

$$\sigma_c = \frac{2K_c N_c P_c R_1^3 Y_a (R_1^3 - R_2^3)}{V_c\left(2K_c R_1^6 Y_a - 2K_c R_1^3 R_2^3 Y_a + 12 R_1^6 \nu_a - 6R_1^6 - 3R_1^3 R_2^3 \nu_a - 3R_1^3 R_2^3\right)}, \qquad (S14)$$

with $Y_a$ being the Young's modulus of the oxidic shell (assumed as the Young's modulus of ilmenite $FeTiO_3 \approx 214$ GPa) and the $\nu_a$ the respective Poisson ratio (assumed as 0.29). Since we consider here the complete phase transformation of the respective FeTiH monohydride, the H/M ratio (indirectly via setting $N_c$) was set to 0.5 for the following calculations.

The total potential energy of the system is given by

$$E = \frac{1}{2}\int_V \varepsilon_{\alpha\beta}\, C_{\alpha\beta\mu\nu}\, \varepsilon_{\mu\nu}\, dV - \int_V \Pi_{\alpha\beta}\, \varepsilon_{\alpha\beta}\, dV. \qquad (S15)$$

Inserting **Equation S7** into **Equation S15** yields an alternative equation for the total potential energy of the system,

$$E = -\frac{1}{2} \int_V^{\square} \Pi_{\alpha\beta} \, S_{\alpha\beta\mu\nu} \, \Pi_{\mu\nu} \, dV + \frac{1}{2} \int_V^{\square} \sigma_{\alpha\beta} \, S_{\alpha\beta\mu\nu} \, \sigma_{\mu\nu} \, dV \ . \tag{S16}$$

Using now the same simplifications as discussed above, i.e., in the derivation of **Equations S7 and S8**, shows that the elastic energies of the core and shell are given by

$$E_c = -\frac{1}{2} \frac{P_c^2 K_c N_c^2}{V_c} + \frac{1}{2} K_c V_c \sigma_c^2 \ , \tag{S17}$$

and

$$E_s = -\frac{1}{2} \frac{P_s^2 K_s N_s^2}{V_s} + \frac{1}{2} \int_{R_1}^{R_2} \sigma_{s,\alpha\alpha} S_{s,\alpha\alpha\mu\mu} \sigma_{s,\mu\mu} \, dV \ . \tag{S18}$$

The last terms of **Equations S17 and S18** are responsible for the microstructure-dependent change of hydrogen´s chemical potential, and summing up the respective last terms and differentiating with respect to $\partial N_c$ yields a term for the microstructure-dependent change in chemical potential, $\Delta\mu$, due to the build-up of elastic backpressure in the FeTi core,

$$\Delta\mu = \frac{\partial}{\partial N_c} \left( \frac{1}{2} K_c V_c \sigma_c^2 + \frac{1}{2} \int_{R1}^{R2} \sigma_{s,\alpha\alpha} S_{s,\alpha\alpha\mu\mu} \sigma_{s,\mu\mu} \, dV \right) \ , \tag{S19}$$

with $dV = 2\pi r^2 \, dr$. Solving this equation yields

$$\Delta\mu = \frac{8 K_c^2 N_c P_c^2 R_1^6 Y_a \left(R_1^3 - R_2^3\right)\left(0.5 K_c V_c Y_a \left(R_1^3 - R_2^3\right) - \pi R_1^3 \left(-4 R_1^3 \nu_a + 2 R_1^3 + R_2^3 \nu_a + R_2^3\right)\right)}{V_c^2 \left(2 K_c R_1^6 Y_a - 2 K_c R_1^3 R_2^3 Y_a + 12 R_1^6 \nu_a - 6 R_1^6 - 3 R_1^3 R_2^3 \nu_a - 3 R_1^3 R_2^3\right)^2} \ . \tag{S20}$$

The subsequent change in the hydrogen pressure in equilibrium with such a hydrogen uptake (namely, an H/M of 0.5, i.e., monohydride formation) is then given by

$$p' = p \exp\left(\frac{2 \, \Delta\mu}{k_b T}\right) \ , \tag{S21}$$

with $p$ being the thermodynamically expected and $p'$ the hydrogen pressures altered by the confinement effect in the core-shell structure.